\documentclass[nofootinbib, amsmath,amssymb, aps]{revtex4}

\usepackage{hyperref}
\usepackage{graphicx}
\usepackage{dcolumn}
\usepackage{bm}
\usepackage{booktabs}
\usepackage{topcapt}
\usepackage{xcolor}
\usepackage{slashed} 
 \usepackage{axodraw2}
 \usepackage{ulem}

\newcommand{\pade}[2]{\begin{small}$[ \, #1 \, | \, #2 \, ]$\end{small}}

\begin{document}

\title{The analytic structure of the Landau gauge quark propagator from Pad\'e analysis}

\author{Alexandre F. Falc\~ao$^{1,2 \, }$}
\email{alexandre.falcao@uib.no}
\author{Orlando Oliveira$^2 \,$}
\email{orlando@uc.pt}
\affiliation{\mbox{} \\
$^1$ Department of Physics and Technology, University of Bergen, 5007 Bergen, Norway \\ \\
$^2$CFisUC, Departament of Physics, University of Coimbra, 3004-516 Coimbra, Portugal \\}

\date{\today}

\begin{abstract}
The analytic structure of the 2 flavour full QCD lattice Landau gauge quark propagator is investigated with Pad\'e approximants 
applied to its vector and scalar form factors.
No poles at complex momentum are observed for the propagator. Moreover, there is clear evidence of a pole at real on-axis negative 
Euclidean momentum, i.e. for Minkowski type of momentum. 
This pole occurs at Euclidean momenta $p^2  \sim - 300$ MeV 
and it reproduces typical quark mass values used in phenomenological effective quark models. 
The Pad\'e approximant analysis also gives hints on the presence of a branch cut. 
Our results also show a clear correlation between the
position of this pole, understood as an effective quark mass,  and the pion mass that is compatible with PCAC.
Slightly differences between the poles for the two quark form factors are observed which can be viewed either as a limitation of the method or 
as a suggestion that the quark propagator has no spectral representation.
\end{abstract}

\maketitle

\tableofcontents

\section{Introduction and Motivation}

The theory that describes the color interaction between quarks and gluons, i.e Quantum Chromodynamics (QCD) (see  \cite{Alkofer:2000wg,Fischer:2006ub,Binosi:2009qm} for reviews),
should also explain hadron phenomenology.  However, so far, in what concerns the experimental evidences for free quarks and gluons \cite{Perl:2009zz,Tanabashi:2018oca} there are no
positive results. This lack of experimental evidences suggests that quarks and gluons can only appear as components of the observed particles, the hadrons, that are 
understood as color singlet objects. This negative results are translated into the confinement hypothesis that still requires to be prooved. 
Despite the success of QCD as a fundamental theory, the bridge linking quarks and gluons to the observed hadrons still needs to be paved. 
A first principle approaches to understand the dynamics of the fundamental quanta of QCD require the knowledge of the fundamental QCD Green functions such as the propagators and vertices, as e.g. the quark-gluon vertex and the three-gluon vertex.

In the functional continuum approach to QCD, the solution of the Dyson-Schwinger equations for the quark, the ghost and the gluon propagators are used to feed Bethe-Salpeter equations, 
Faddeev equations, etc. to access the properties of the physical hadronic states. The computation of the solutions of these equations requires the knowledge of the propagators in the complex 
momentum plane. Therefore, the analytic structure of the propagators have importante roles.
Indeed, being able to identify the poles and branch cuts of the propagators is crucial to the bound state problem 
but also for the comprehension of the confinement mechanism, to the dynamical properties of quarks and gluons and, ultimately  to the understanding of QCD itself.

For the gluon and ghost sectors of QCD, the past twenty years saw an effort to have reliable descriptions and interpretations for the propagators. A recent review
on the gluon propagator can be found in \cite{Papavassiliou:2022wrb}. See,  also,
\cite{Leinweber:1998uu,Becirevic:1999uc,Becirevic:1999hj,Cucchieri:2007md,Bogolubsky:2009dc,Dudal:2010tf,Cucchieri:2011ig,Maas:2011ez,Dudal:2018cli,Li:2019hyv,Catumba:2021hcx}
and references therein for lattice calculations and 
\cite{vonSmekal:1997ohs,Alkofer:2003jj,Boucaud:2007va,Boucaud:2007hy,Huber:2007kc,Dudal:2008rm,Aguilar:2008xm,Rodriguez-Quintero:2010qad,Aguilar:2010zx,Strauss:2012zz,Eichmann:2021zuv,Pelaez:2014mxa,Aguilar:2019uob,Reinosa:2020skx,Fischer:2020xnb,Pelaez:2021tpq,Dudal:2022nnu,
Morris:2005tv,Morris:2006in,Arnone:2005fb,Zwanziger:2012xg,Allendes:2014fua,Meyers:2014iwa,Siringo:2014lva,
Dudal:2015khv,Cyrol:2016tym,Gogokhia:2016rix,Cyrol:2017ewj,Cyrol:2018xeq,Lowdon:2018mbn,Siringo:2018uho,Mintz:2018hhx,Aguilar:2020uqw,Huber:2020keu,Napetschnig:2021ria,Horak:2021syv,Aguilar:2021uwa,Gracey:2022xsz,Horak:2022aqx,Horak:2022myj,Aguilar:2022wsh}
and references therein for continuum approaches. 
On the other hand, the ghost propagator has also been thoroughly studied, as can be found in 
e.g. \cite{vonSmekal:1997ohs,Alkofer:2003jr,Aguilar:2008xm,Bogolubsky:2009dc,Duarte:2016iko,Boucaud:2017ksi,Duarte:2017wte} and references therein.
Similarly, one can find in the literature several studies for the quark propagator, see e.g 
\cite{Krein:1990sf,Bowman:2002bm,Boucaud:2003dx,Fischer:2005nf,Furui:2006rx,Furui:2006ks,Ayala:2012pb,
Burgio:2012ph,Dorkin:2013rsa,Dorkin:2014lxa,Fu:2015tdu,Windisch:2016iud,
Oliveira:2018lln,Solis:2019fzm,Lessa:2022wqc,
Comitini:2021kxj,Virgili:2022wfx}, that include continuum and lattice
QCD calculations. From the good agreement obtained by several groups that use different techniques, it seems reasonable to claim a fair understanding of the QCD two point correlation functions. 
However, most non-perturbative computations are done using the Euclidean formulation of the theory, which limits the determination of the analytical structure of the propagators.

The quark propagator has been investigated by several groups
\cite{Maris:1994ux,Maris:1995ns,Burden:1997ja,Bhagwat:2003vw,Alkofer:2003jj,Frederico:2019noo,Hayashi:2021nnj,Hayashi:2021jju,Sauli:2020dmx}, with the Dyson-Schwinger studies suggesting the 
presence of poles at complex momenta. However, it is difficult to disentangle if the complex poles and analytic structure result directly from the truncations and/or the parameterization of the vertices
used to solve the Dyson-Schwinger equations themselves. 
Therefore, it is important to investigate the analytical structure of the QCD propagators with different methods, different sets of data, and to confront the outcome of  different techniques.

The main goal of the current work is to address the analytic structure of the quark propagator by studying a subset of the Landau gauge lattice data generated in \cite{Oliveira:2018lln}, by
applying the techniques based on Pad\'e approximants considered in \cite{Falcao:2020vyr} for the pure Yang-Mills gluon and ghost propagators.

\begin{figure}[t]
   \centering
   \includegraphics[scale=0.31]{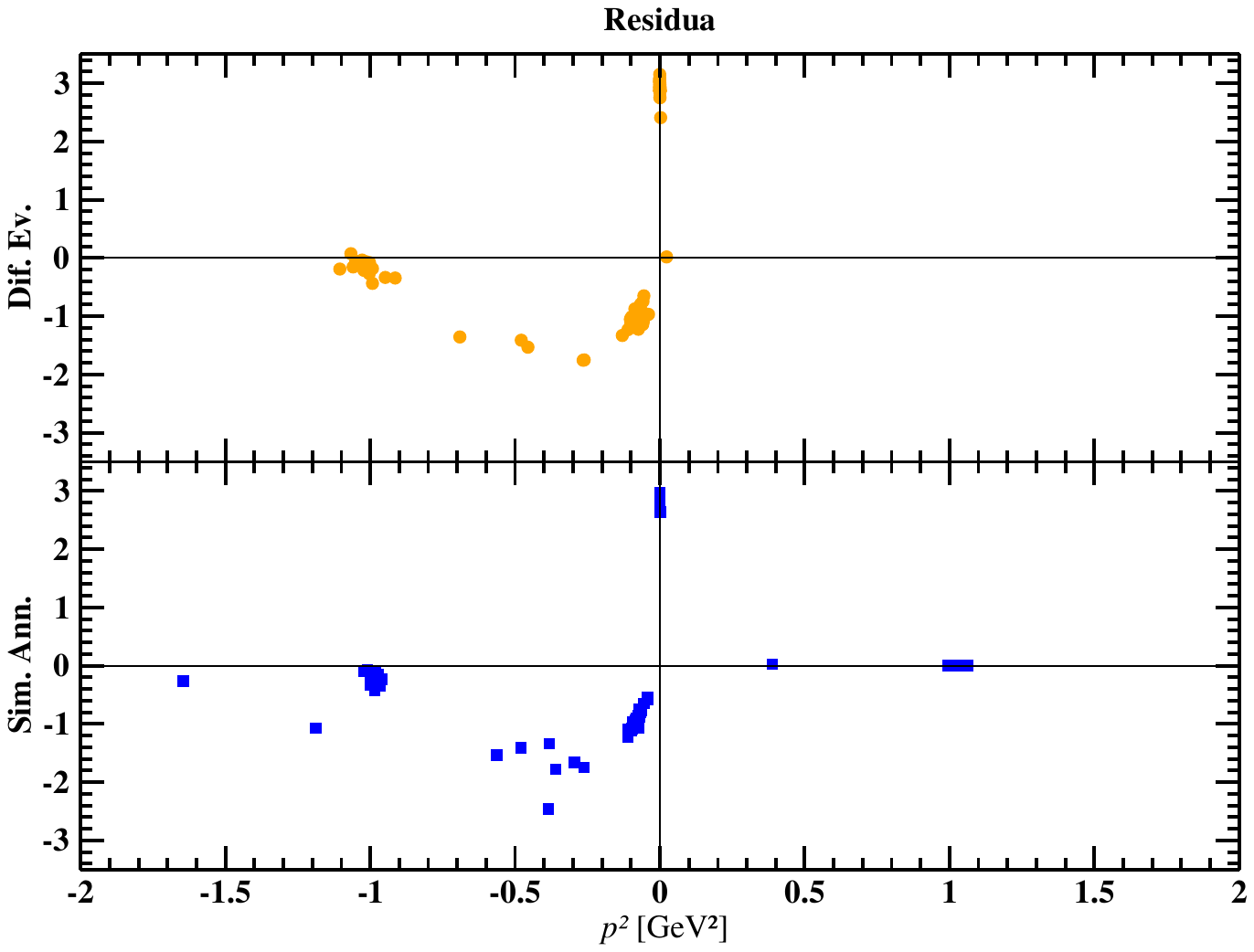}  
   \includegraphics[scale=0.31]{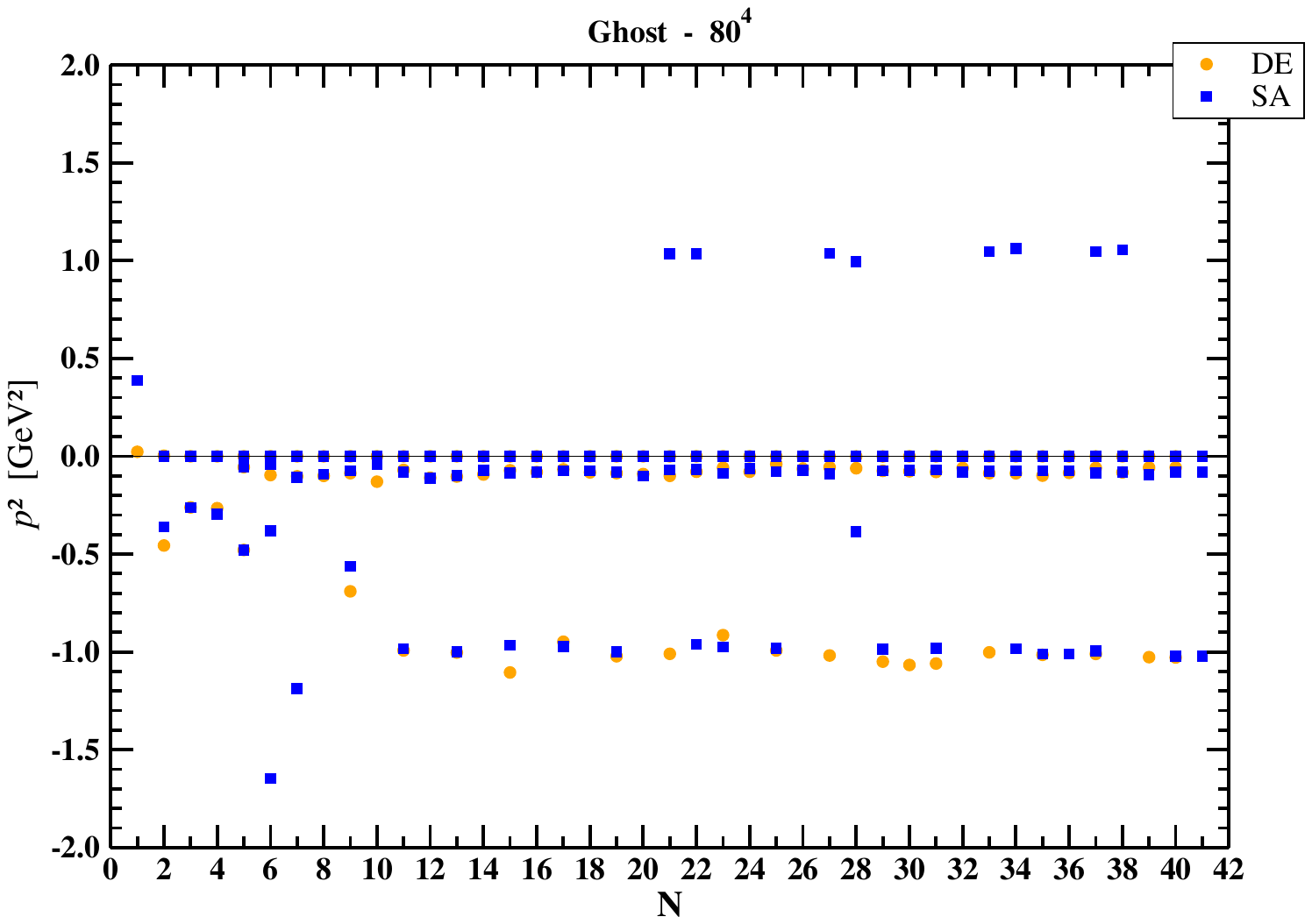} 
   \caption{Pad\'e estimations for the ghost propagator residua for real on-axis momenta as a function of $p^2$ (left) and a function of the degree of the Pad\'e approximant (right).}
   \label{fig:ghost}
\end{figure}

In this previous work \cite{Falcao:2020vyr}, the study of the analytic structure of the pure Yang-Mills theory gluon and ghost propagators used only the absolute value of the residua and  
their relative importance. Herein, we take the opportunity to complement this information and look also at the sign of the relevant residua. 
The sign of the residua provides a naive and clear interpretation of the corresponding poles. In this respect, for real on-axis momenta, see Fig. \ref{fig:ghost}, some of the ghost residua are 
negative and the poles cannot represent physical particles. For the gluon propagator and for real on-axis momenta, the interpretation of the residua is not so clear, see Fig. \ref{fig:gluon128}, and
no conclusions can be established firmly. 
As discussed in \cite{Falcao:2020vyr}, the Pad\'e analysis for the gluon propagator identifies clearly a pair of complex conjugate poles that are 
not seen in the ghost propagator.   In \cite{Boito:2022rad}  the analytical structure of the gluon and ghost propagators using SU(2)  Landau gauge lattice data was investigated with Pad\'e approximants and the authors come to similar conclusions to those found in \cite{Falcao:2020vyr}. 

The analysis of quark propagator data with Pad\'e approximants discussed below shows an analytical structure that differs from the 
analytic structure of the gluon and the ghost propagators. Contrary to the Dyson-Schwinger studies for complex momenta, no evidence is found for poles at complex momenta. 
On the other hand, the Pad\'e analysis  gives indications of a possible branch cut for real on-axis momenta.   Furthermore,
the Pad\'e approximant description of the lattice data finds poles at real on-axis negative Euclidean momenta around $p^2 \sim - 300$ MeV, i.e. for Minkowski momenta, that are correlated with the pion mass.
Indeed, it is observed that the correlation between the quark mass, understood as the pole mass at the Minkowski momenta, and the pion mass is compatible with the predictions of PCAC. 
Moreover, the residua of these poles for time-like Euclidean momenta is positive. This seems to indicate that the quark propagator has a particle-like pole at Minkowski momenta, at momenta that approximately 
reproduce typical values for  effective quark masses, and further structures, that we are not able to disentangle, that must prevent the quark from being a physical particle. 
Recall that, for example, the Dyson-Schwinger studies for the quark propagator observe positivity violation. Positivity violation implies that if a 
K\'all\"en-Lehmann representation of the propagator exists\footnote{
Strictly speaking the K\'all\"en-Lehmann representation of a two-point function is an integral representation for this correlation function that introduces
a positive defined spectral density with the usual meaning found in QFT textbooks. When, in the integral representation of the propagator,
the would-be spectral density ceases to be positive over all its domain, we no longer have a K\'all\"en-Lehmann representation but, instead, an integral
representation of the K\'all\"en-Lehmann type. Herein, we follow the usual notation and use the name spectral density or spectral representation in
any case.}, then its 
spectral representation is not positive defined. A spectral  representation
that is not positive defined is not compatible with a quantum mechanical probabilistic interpretation and single quark state cannot belong to the 
set of physical states of QCD. 
Naively one could think that the presence of the pole at Minkowsky type of momenta with a positive residua implies that such type of pole would 
correspond to a physical particle but this it not necessarily true. A counter example can be find in \cite{Solis:2019fzm}, where the authors provide a quark propagator built from a toy model 
that have similar properties as those just described but do not represent a physical particle. 
The pole at the Minkowsky momenta occurs for momenta that are within typicall values for the quark mass used in effective quark models. In this sense, the results found herein
seem to give some support to the effective quark models and to the phenomenological values that they use for the quark mass. 
A clarification of these questions calls for an understanding of what are the signs of confinement in the quark propagator, a complex problem that clear goes beyond the type of answers that we are able
to provide in the current work. 

The manuscript is organised as follows.
In Sec. \ref{Sec:Pade} the Pad\'e approximant theory to address the analytical structure of the propagators is reviewed.
Then, the results for the pure Yang-Mills theory for the ghost and the gluon propagators are revisited in Sec. \ref{Sec:ghost} and Sec. \ref{Sec:gluon}, respectively. 
%
The analysis of quark propagator data with Pad\'e approximants is performed in Sec. \ref{Sec:quarkprop}. 
Finally, in Sec. \ref{Sec:summary} we summarise the results and conclude.

\section{Pad\'e approximants \label{Sec:Pade}}

The Pad\'e approximant to a function $f(x)$ defined in the interval $x \in [a , b]$ looks for the 
best approximation to $f(x)$ written as a ratio of polynomials 
in $x$. The polynomials associate zeros and poles to $f(x)$, corresponding to the zeros of the numerator and denominator, respectively. The location of these zeros and
poles depends on the degree of the polynomials considered and changing the degrees of the numerator and denominator polynomals also changes the position of the zeros and the poles.
However, in general, there is a subset of zeros and poles whose position does not depend on the Pad\'e approximant used. These zeros and poles, which 
are independent of the Pad\'e approximant, can  be associated with the analytic structure of $f(x)$. The remaining zeros and poles are
artefacts of the approximation. Moreover, 
a stable structure of in the complex plane formed by a sequence of alternating poles and zeros of the approximant can represent a branch cut.

For particle propagators in Quantum Field Theory represented, generally, as $D(p^2)$, the Pad\'e approximant of order \pade{M}{N} to the propagator
is given by
\begin{equation}
   D(p^2) \approx P^M_N(p^2) = \frac{Q_M(p^2)}{R_N(p^2)} \ ,
   \label{Eq:usual_Pade_M_N_0}
\end{equation}
with
\begin{eqnarray}
     Q_M(p^2)  =  q_0 + \cdots + q_M \, \left( p^2 \right)^M \qquad\mbox{ and }\qquad
     R_N(p^2)  =  1 + \cdots + r_N \, \left( p^2 \right)^N 
\end{eqnarray} 
and where, by convention, the coefficient of the lowest order term in the denominator is set to one. The problem of building a Pad\'e approximant to $f(x)$
is reduced to find the set of coefficients $q_0, \, q_1, \dots, \, r_1, \dots$ that define the best approximation to $D(p^2)$ according to Eq. (\ref{Eq:usual_Pade_M_N_0}).
Pommerenke’s Theorem \cite{Pomm73} states that for a meromorphic function $f(z)$, the Pad\'e sequences \pade{M}{M+k}, with fixed $k$, converge to $f(z)$ 
in any compact set of the complex plane.
This theorem gives support to use the Pad\'e approximant sequences in the investigation of the analytic structure of particle propagators.

 In the sequences of Pad\'e approximants \pade{M}{M+k}, single poles of $f(z)$ appear as stable poles for sufficiently large values of $M$. 
On the other hand, the Froissart doublets \cite{Baker75,BaMo96,BeOr99,SanzCillero:2010mp,Queralt:2010sv} are those poles that depend strongly on the degree 
of the polynomials or have nearby zeros. The presence of the nearby zeros for the Froissart doublets result in residua whose absolute value is small.
The absolute value of the residua can then be used to distinguish the meaningful poles from those that are artefacts of the method. 
The doublets that appear at sufficiently large values of $M$ are artefacts associated with the use of ratio of polynomials to represent $f(x)$.

It is common practice to use diagonal  \pade{M}{M} or near diagonal \pade{M}{M \pm 1} Pad\'e approximants in the representation of $f(x)$. In our analysis of the quark propagator,
motivated by the studied of the pure Yang-Mills gluon and ghost propagators \cite{Falcao:2020vyr}, and also to be closer to the perturbative description of the propagator,
that should be recovered at high momenta, only the sequence \pade{N}{N +1} will be considered.

For the determination of the coefficients of the polynomials that define a given Pad\'e approximant  we look at the absolute minimum of the objective function 
\begin{equation}
   \chi^2 = \sum_{j=1}^{N_{mom}} \left( \frac{ D(p^2_j) - D_{Lat}(p^2_j) }{ \sigma(p^2_j) } \right)^2 \ ,
   \label{Eq:objectivefunction}
\end{equation}
where the sum is over the data points obtained with lattice QCD simulations, i.e. the number of lattice momenta $N_{mom}$ accessed in the simulation, 
$D(p^2) = P^N_{N+1}(p^2)$, $D_{Lat}(p^2)$ are the data points for the given function and $\sigma(p^2)$ are the associated statistical
error with $D_{Lat}(p^2)$. In the following we do not take into consideration correlations between the momenta. In this way, the numerical problem becomes a non-linear global optimisation problem that is
solved with the help of the global optimisation methods available within \textit{Mathematica}  \cite{Math} software package, i.e. the \textit{Mathematica} implementations of the differential evolution (DE) method and of the 
simulated annealing (SA) method. We recall the reader that global optimization problems are non-trivial \textit{per se} and the comparison of the two methods is important to have confidence in the
results. Furthermore, we report that in \cite{Falcao:2020vyr} the authors performed a series of tests to understand the numerical performance of the numerical method before studying the 
analytic structure of the gluon and ghost pure Yang-Mills propagators. We refer the interested reader to this work for further details concerning numerical experiments, 
performance and limitations of the method.

\section{Resum\'e of the quenched analysis}

As a warm up to the analysis of the Landau gauge full QCD quark propagator, we revisit 
the results of  Pad\'e approximants analysis for the analytic structure of the pure gauge gluon and ghost propagators
performed in \cite{Falcao:2020vyr}, taking the opportunity to comment on the value of the residua.

\begin{figure}[t]
   \centering
   \includegraphics[scale=0.31]{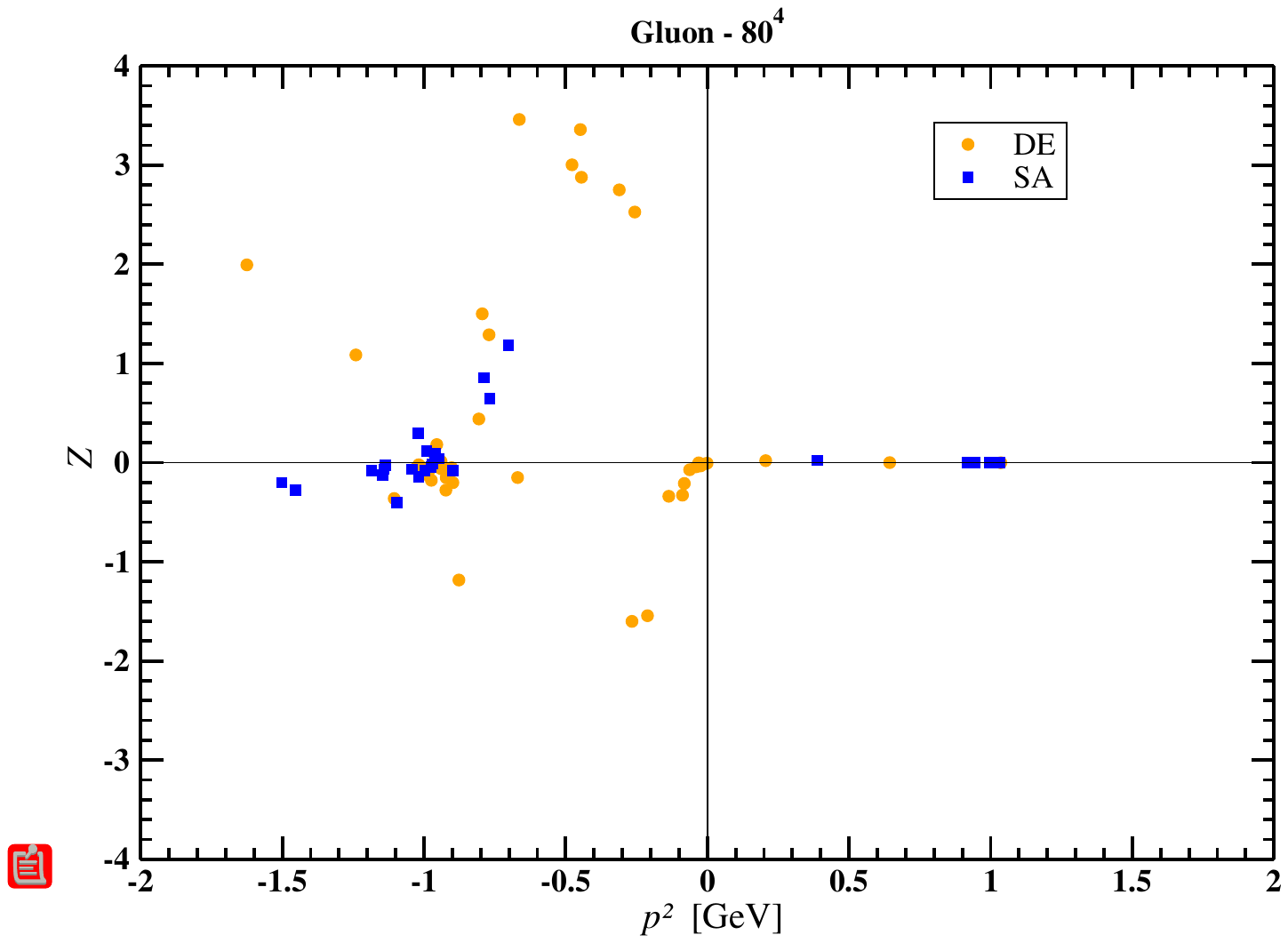}  
   \includegraphics[scale=0.31]{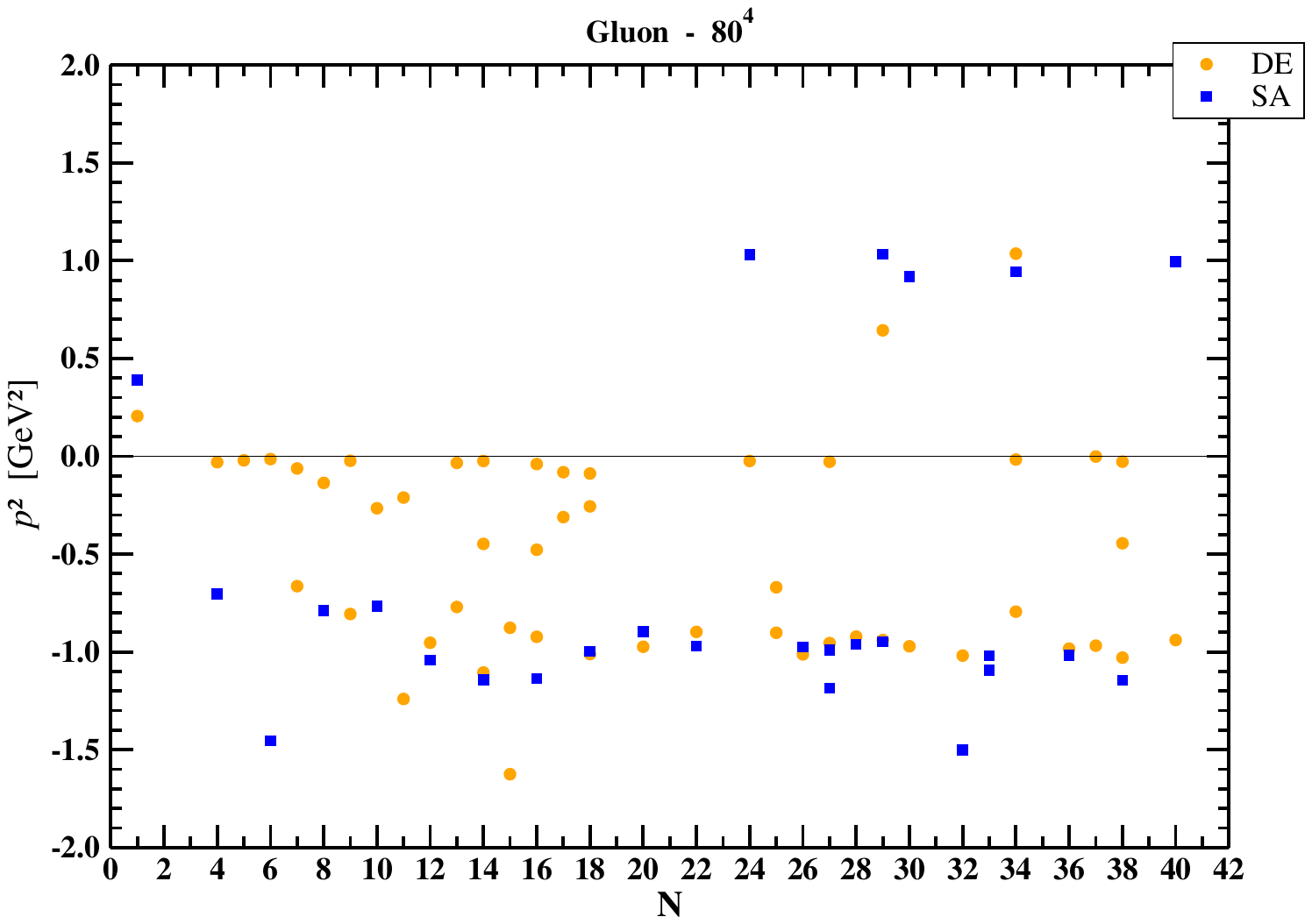} \\ 
%
   \includegraphics[scale=0.31]{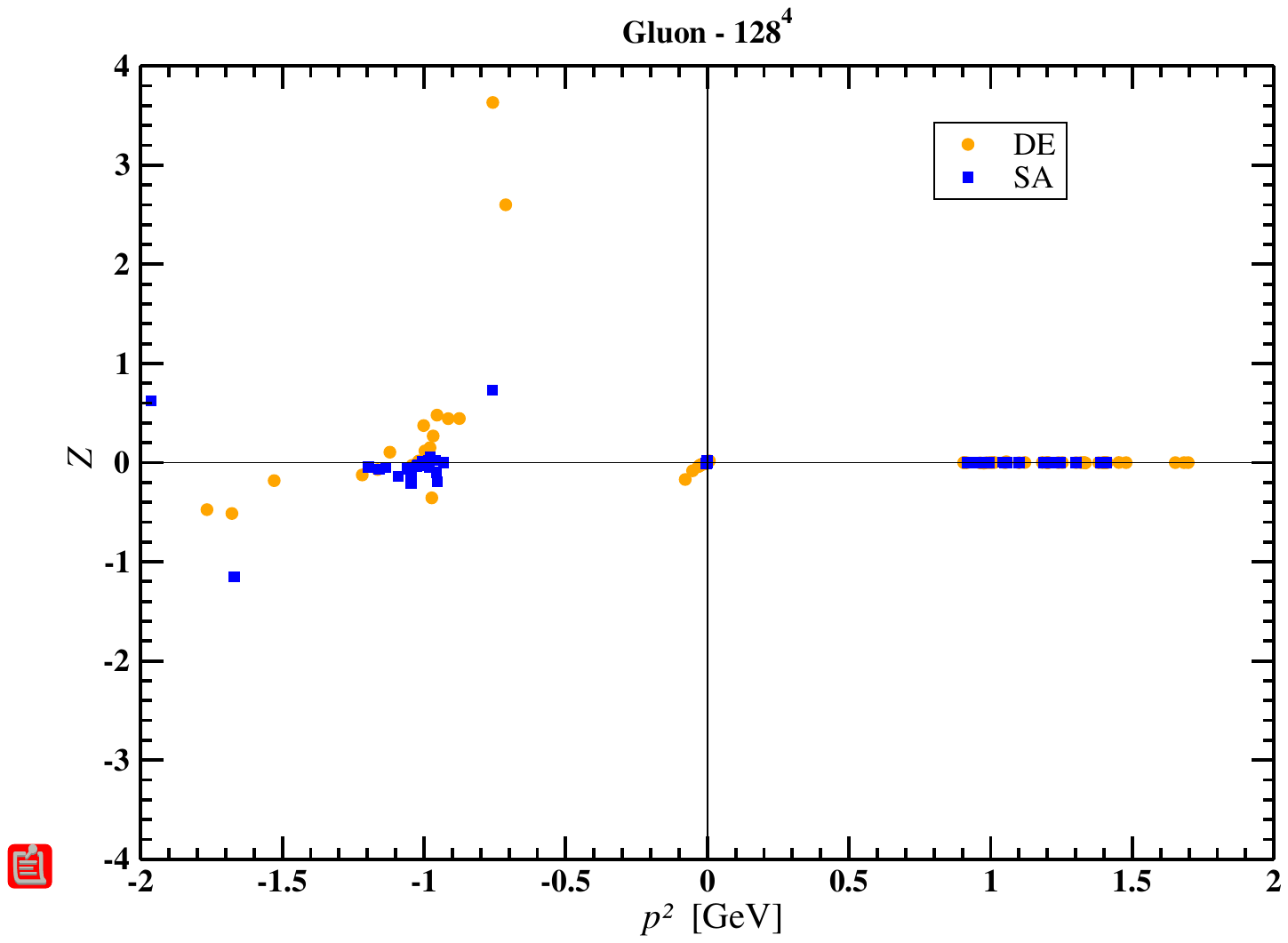}  
   \includegraphics[scale=0.31]{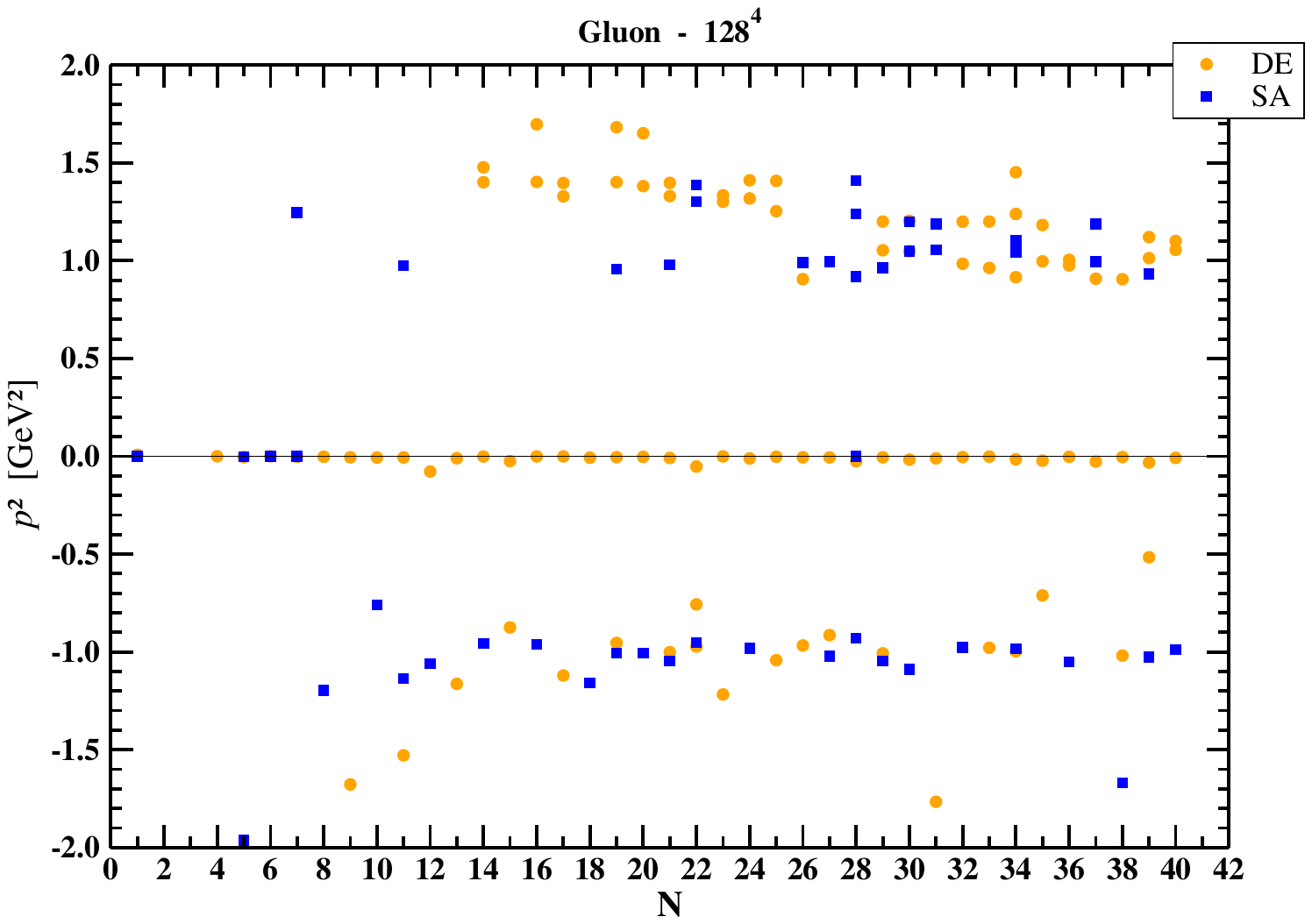} 
   \caption{Pad\'e estimations for the gluon propagator residua for real on-axis momenta for the simulations using the two largest lattice volumes considered in \cite{Falcao:2020vyr}.}
   \label{fig:gluon128}
\end{figure}

\subsection{The Ghost Propagator \label{Sec:ghost}}

The analysis of the Landau gauge ghost propagator analytic structure 
concluded for the absence of poles at complex momenta, for the presence of  a pole at zero momenta and 
found possible evidences for a branch cut at time-like (Minkowski) momenta. 
The clear identification of the branch points is not well established and it requires further studies,
that call for simulations with larger ensembles of gauge configurations and/or larger physical volumes.  
The analysis of the analytic structure of the propagator in \cite{Falcao:2020vyr}
was centred on the absolute value and on the relative importance of the absolute value of the residua. 

The Pad\'e estimations of the residua $Z$ for the ghost propagator for real on-axis monenta are reported in Fig. \ref{fig:ghost}. 
For real on-axis momenta, the residua are all real numbers as required if the propagator is a real function of $p^2$. 
The data shows clearly that the largest residua has $Z \sim 3$ and is associated with the pole at zero momentum.
This pole appears for all the $N$ considered in the analysis. 
The second highest residuum in absolute value is negative and, therefore, cannot represent a physical particle.
Furthermore, it is associated with a negative $p^2$ that corresponds to Minkowsky or time-like momenta. 
As seen in Fig. \ref{fig:ghost}, there is a pole at $p^2 \sim -0.5$ GeV$^2$ that both methods identify only for small $N$,
and another pole at $p^2 = -1$ GeV$^2$ also with a negative residuum that is identified for $N \geqslant 11$ and 
whose position seems to be independent of the degree of the approximant\footnote{ For $N = 11$ the differential evolution method identifies a pole at $p^2 = -0.992$ GeV$^2$ with a 
residuum of $Z = -0.4360$, while the simulated annenaling sees a pole at $p^2 =-0.984$ GeV$^2$ with $Z = -0.4307$.}.
As discussed in \cite{Falcao:2020vyr}, see their Fig. 18, the pole associated with $p^2 \sim -0.5$ and $-1$ Gev$^2$ have
an associated close by zero of the numerator that suggests the presence of a branch point for Minkowski momenta. 
As observed in  the above cited work, there is another pole at smaller momentum,  that is stable under variation
of degree of the Pad\'e approximant, whose residua is negative and with no close by zero of the polynomial
approximation\footnote{For the differential evolution this pole appears firstly for $N = 6$ and for $p^2 = -0.096$ GeV$^2$ with a $Z =-1.006$,
while the simulated annealing returns for $N = 6$ a $p^2 = -0.042$ Gev$^2$ for a $Z = -0.548$.}. 
Its residuum being real and negative, the pole cannot represent a physical particle. 
The remaining observed poles have much smaller $|Z|$, their position is not stable against the variation of $N$ and, therefore, we do not consider them as having a physical meaning.
We call the reader's attention that the analysis of the infrared region ($p^2 < 1$ GeV$^2$) for the same ghost propagator data performed in \cite{Li:2021wol},
although in a particular theoretical framework, 
also encountered a dominant pole at $p^2 = 0$ with a positive residuum and a second pole at $p^2 \sim -0.3$ GeV$^2$
with a negative residuum but with smaller absolute value.

\subsection{The Gluon Propagator \label{Sec:gluon}}

The Pad\'e analysis of the Landau gauge lattice gluon propagator identifies a pair of complex conjugate poles at 
$p^2 \sim - 0.28 \pm \, i \, 0.43$ GeV$^2$. No poles or branch cuts are clearly identified. 
Indeed, as discussed in \cite{Falcao:2020vyr}, there are indications of a possible branch cut but, as for the ghost propagator,
no firm conclusions were drawn. 

The residua for the gluon propagator estimated using Pad\'e approximants for on-axis momenta are all real numbers and 
are reported in Fig. \ref{fig:gluon128} for the two larger lattice volumes considered previously and using the two
global optimization methods, i.e. differential evolution and simulated annealing. As can be observed in Fig. \ref{fig:gluon128} 
there seems to be stable poles at $p^2 \sim \pm 1$ GeV$^2$ but their residua are small in absolute value and, for some
of the $N$, they have close by zeros; see Fig. 14 in \cite{Falcao:2020vyr}. These two features prevent us to
assign a meaning to these poles besides being possible branch points. Note also that, for on-axis momenta, the two global optimization methods 
identify structures in the same range of momenta but their results are not fully consistent.


\section{The Quark Propagator \label{Sec:quarkprop}}

Let us now look at the analysis of the Landau gauge full quark propagator for a subset of the data published in
\cite{Oliveira:2018lln}. In the following, we will investigate  the results of the ensembles simulated at $\beta = 5.29$, 
that corresponds to a lattice spacing of $a = 0.071$ fm, and from the available data only the data for pion masses of 
422 MeV, 290 MeV and 150 MeV.  The first simulation was performed on an asymmetric $32^3\times64$ lattice, 
while the last two simulations used a $64^4$ lattice. The simulation with an almost physical pion mass uses about
half of the gauge configurations of the other two cases and, therefore, the associated statistical errors are  larger. 

\begin{figure}[t]
   \centering
   \includegraphics[scale=0.31]{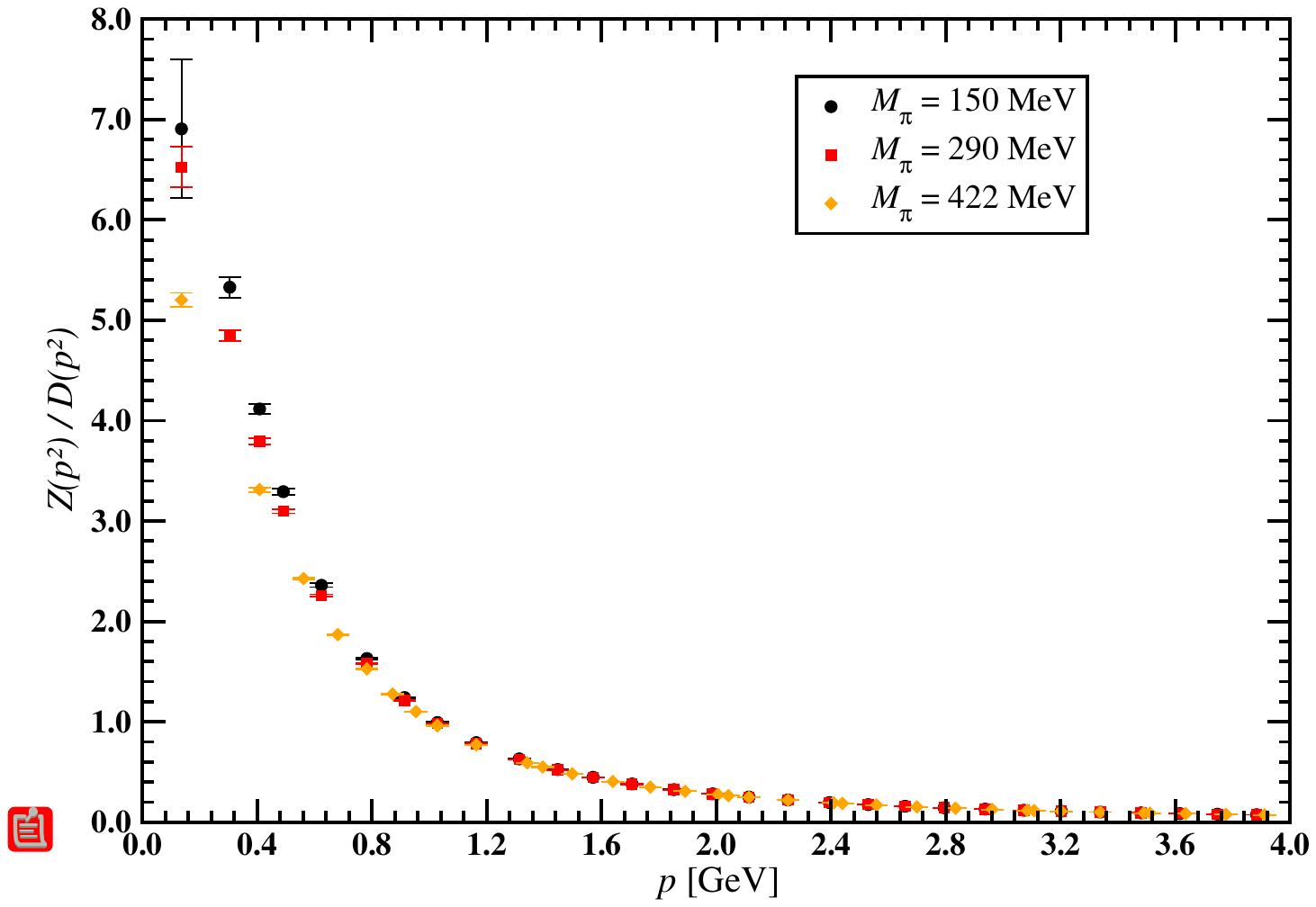}  
   \includegraphics[scale=0.31]{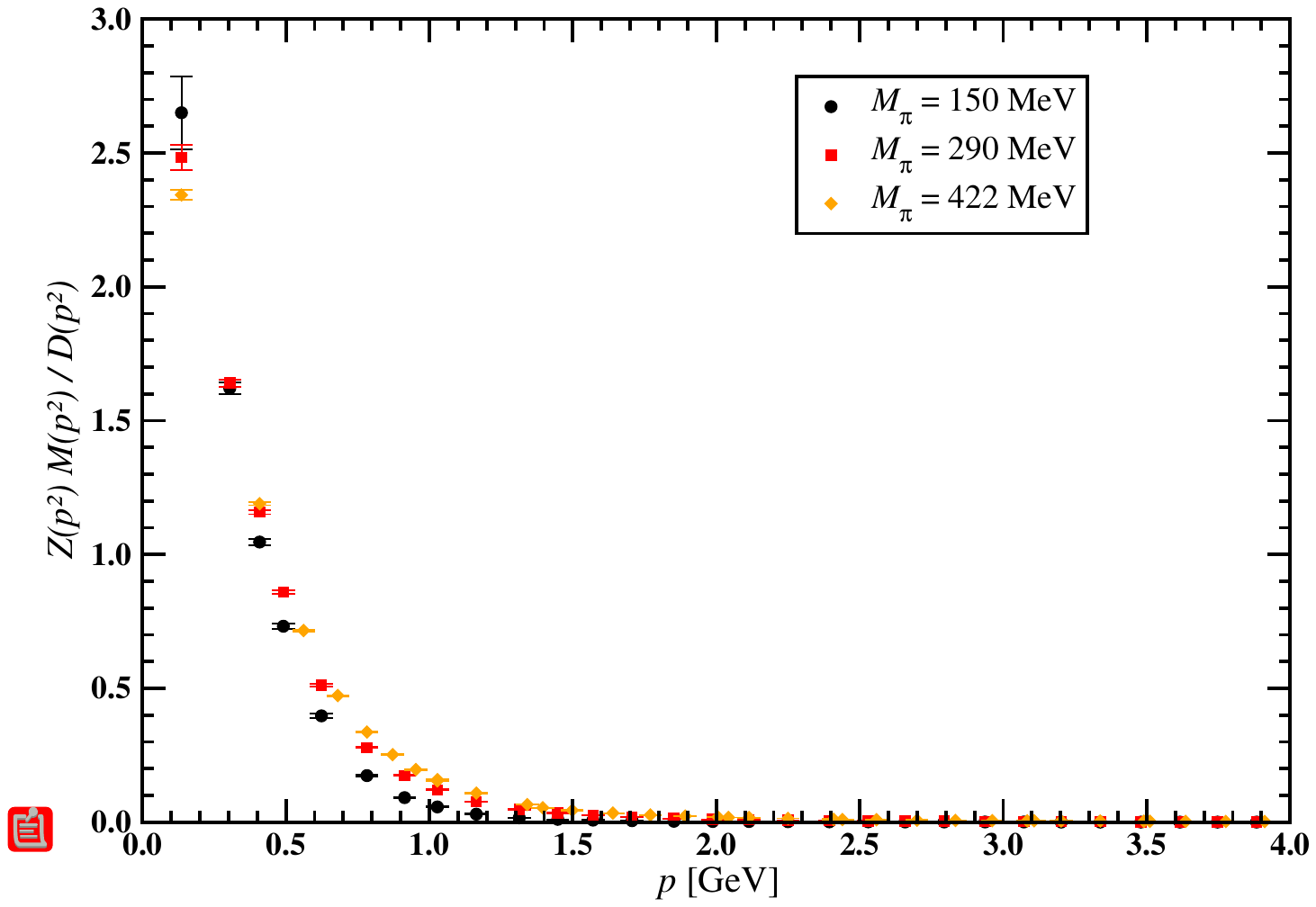}  
   \caption{Lattice Landau gauge quark propagator data for the vector (left) and scalar (right) form factors. See text for definitions.}
   \label{fig:quarkdata}
\end{figure}

\begin{figure}[t]
   \centering
   \includegraphics[scale=0.25]{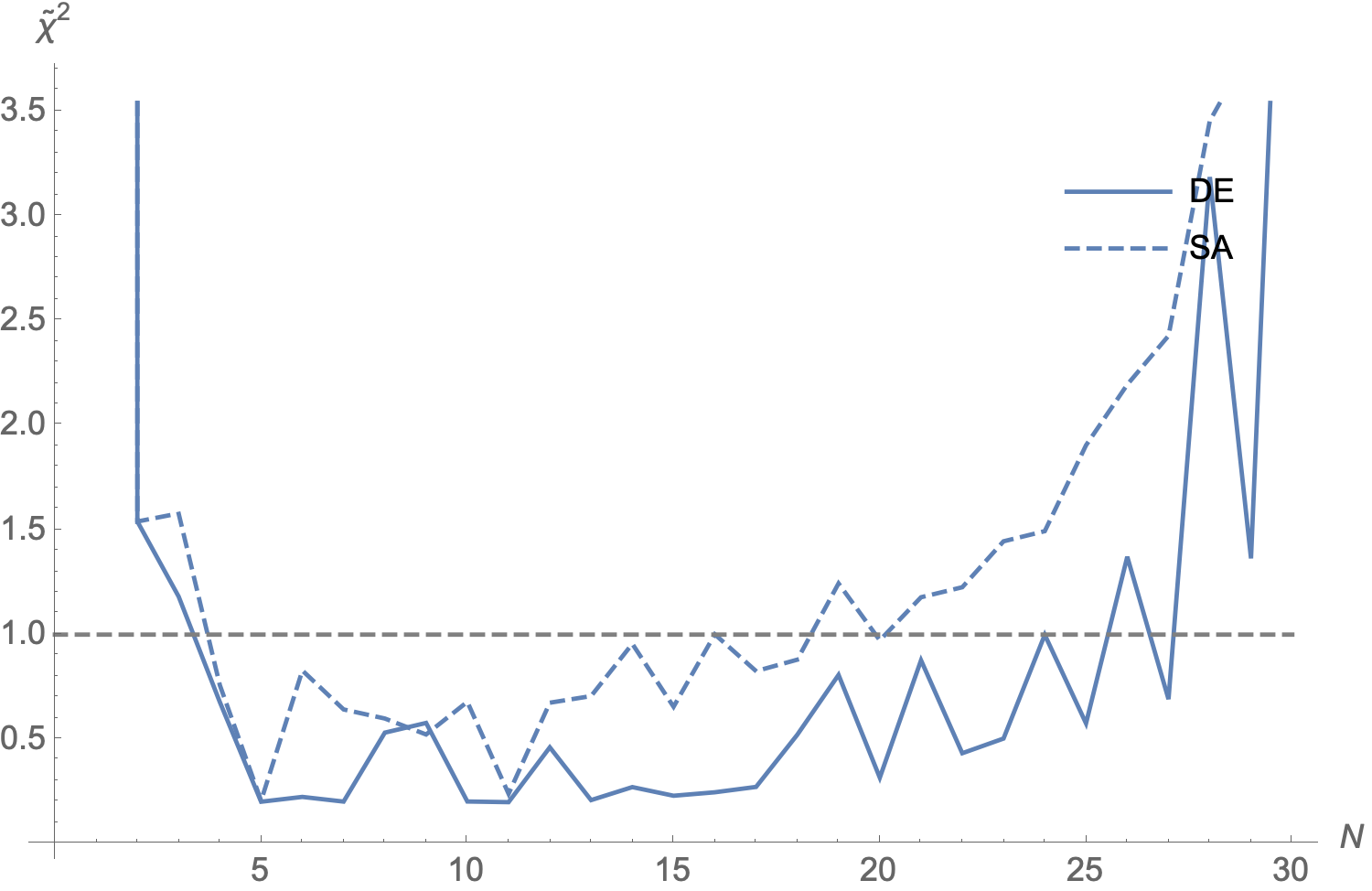}  \hspace{0.75cm}  
   \includegraphics[scale=0.25]{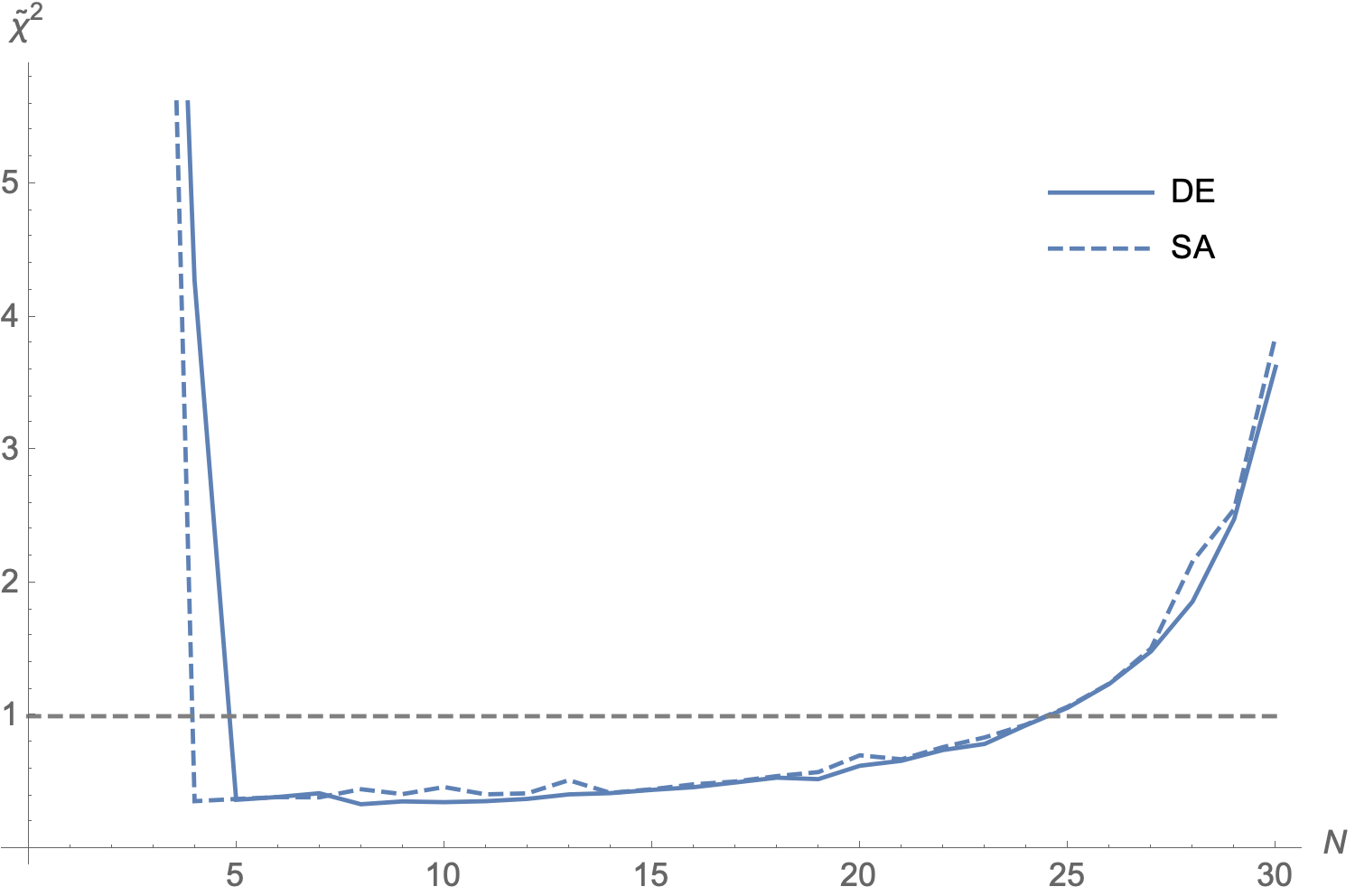}  \\
   \includegraphics[scale=0.25]{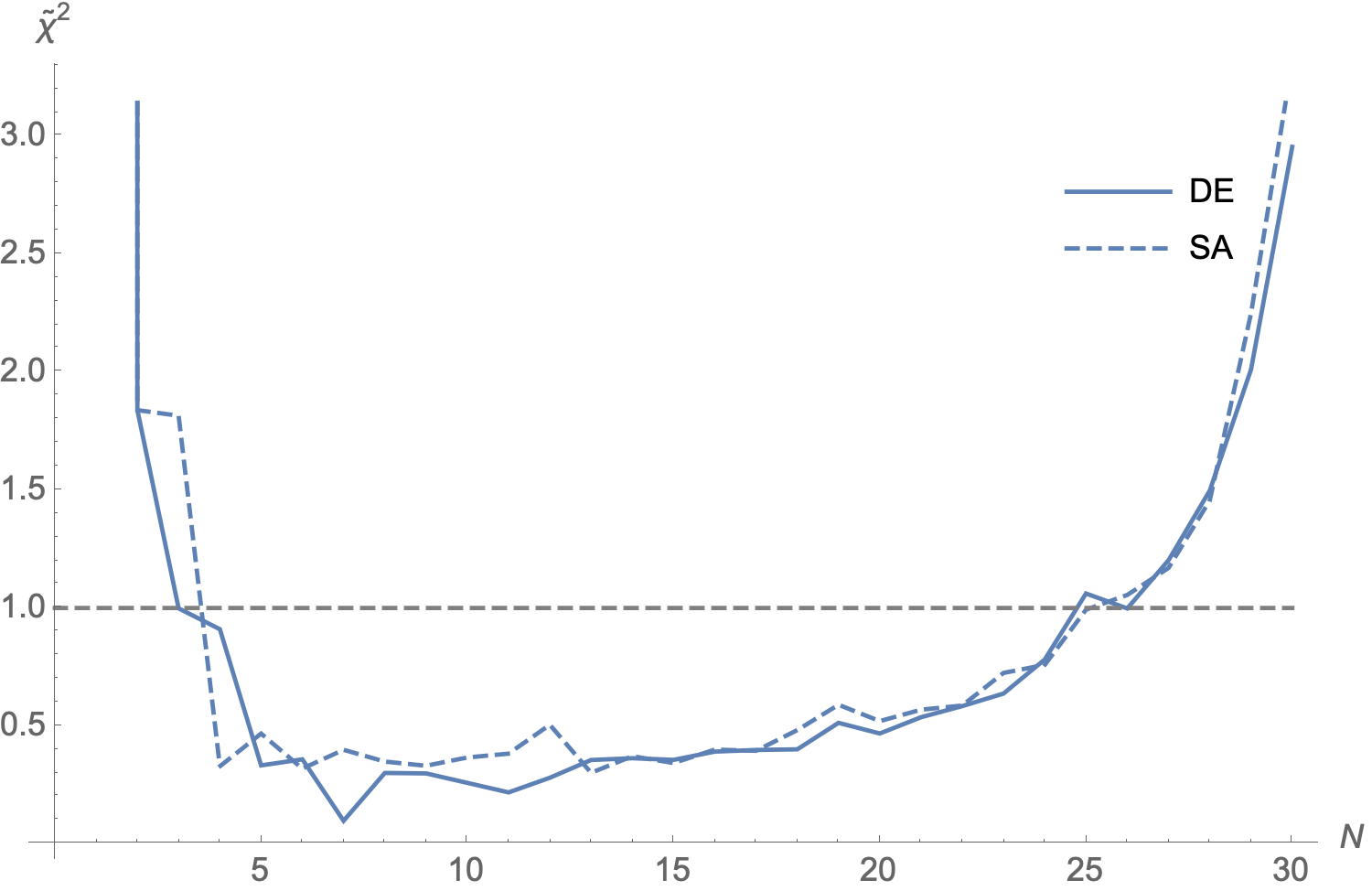}  \hspace{0.75cm}  
   \includegraphics[scale=0.25]{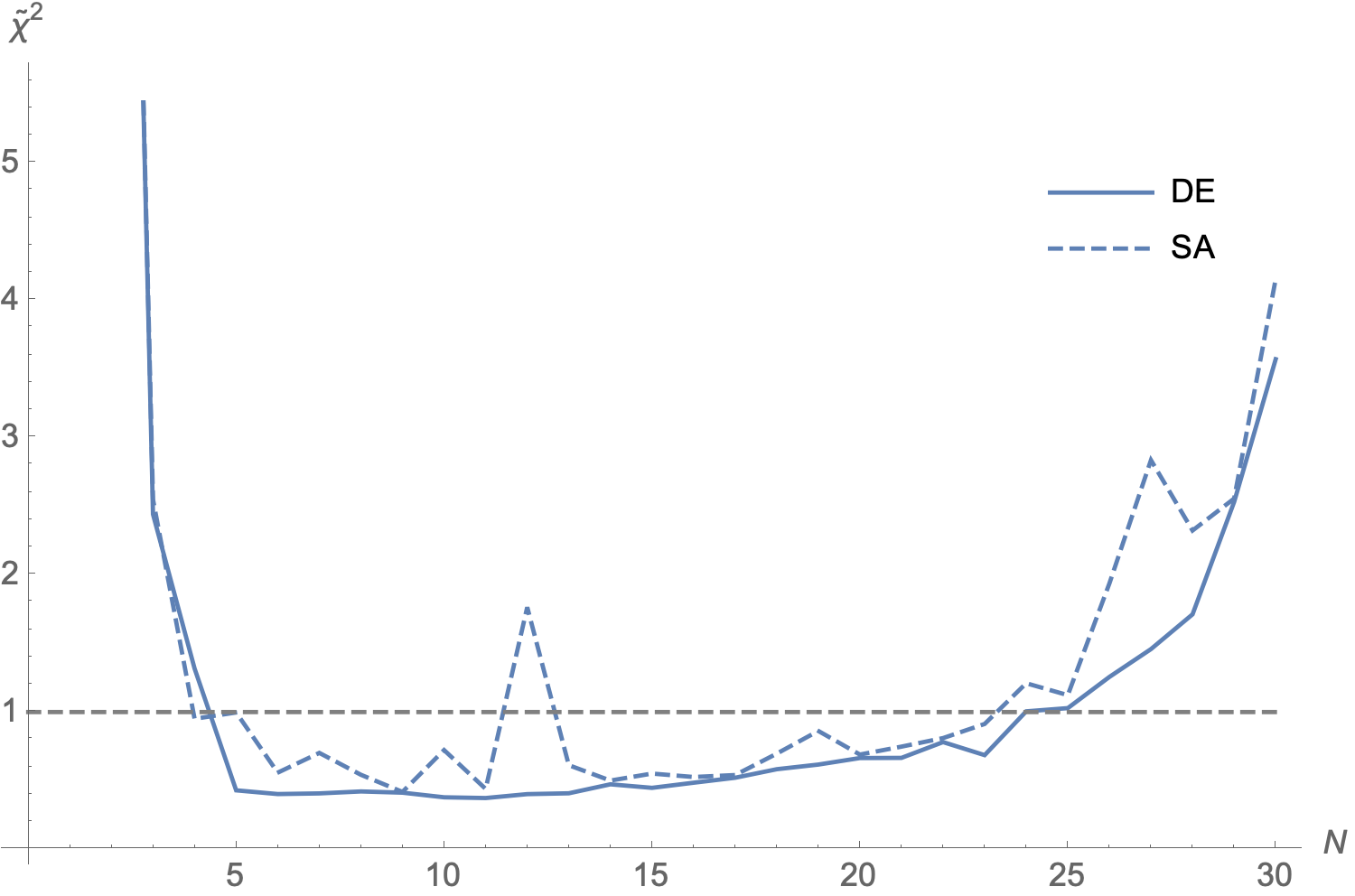}  \\
   \includegraphics[scale=0.25]{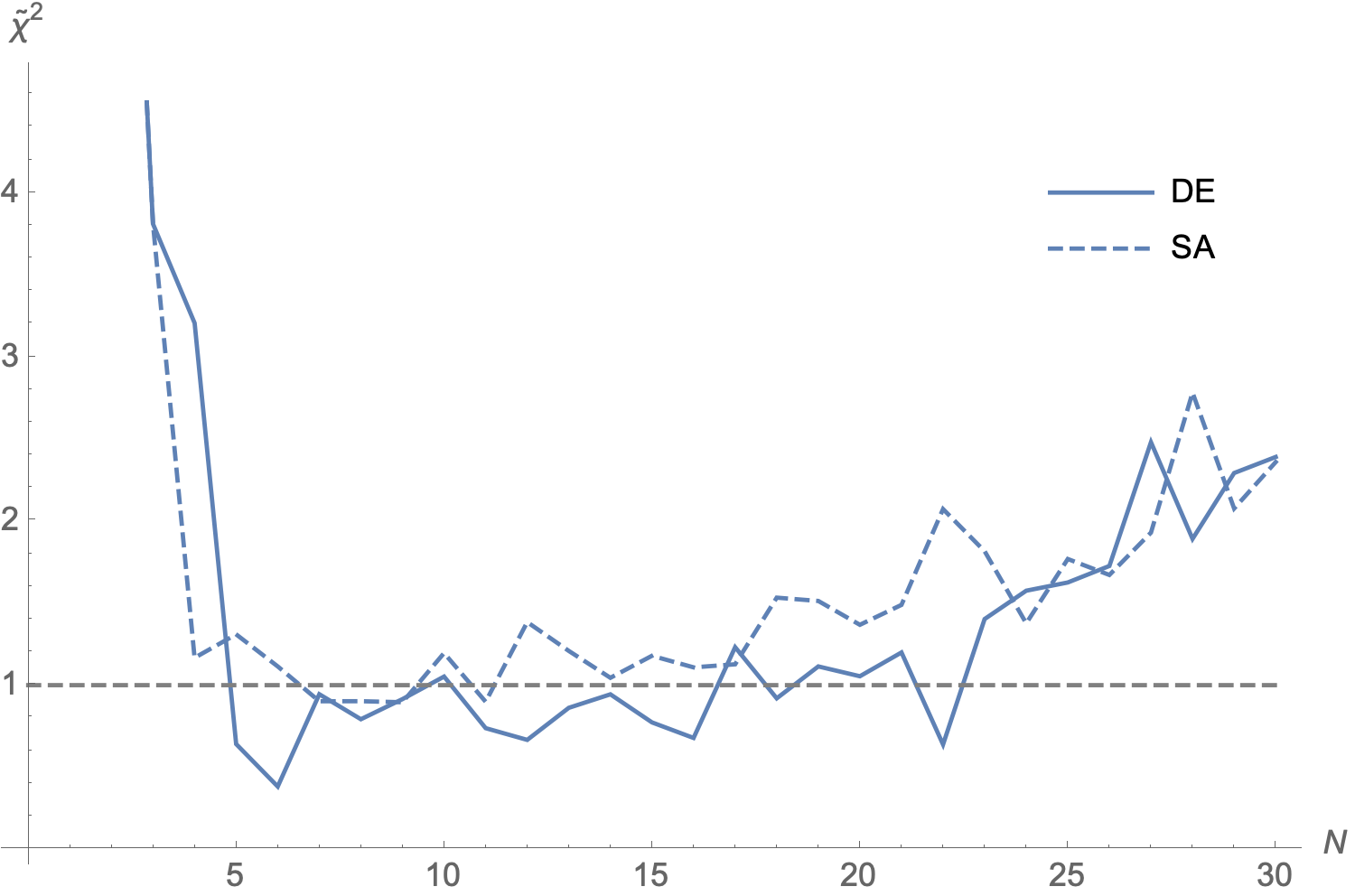}  \hspace{0.75cm}  
   \includegraphics[scale=0.25]{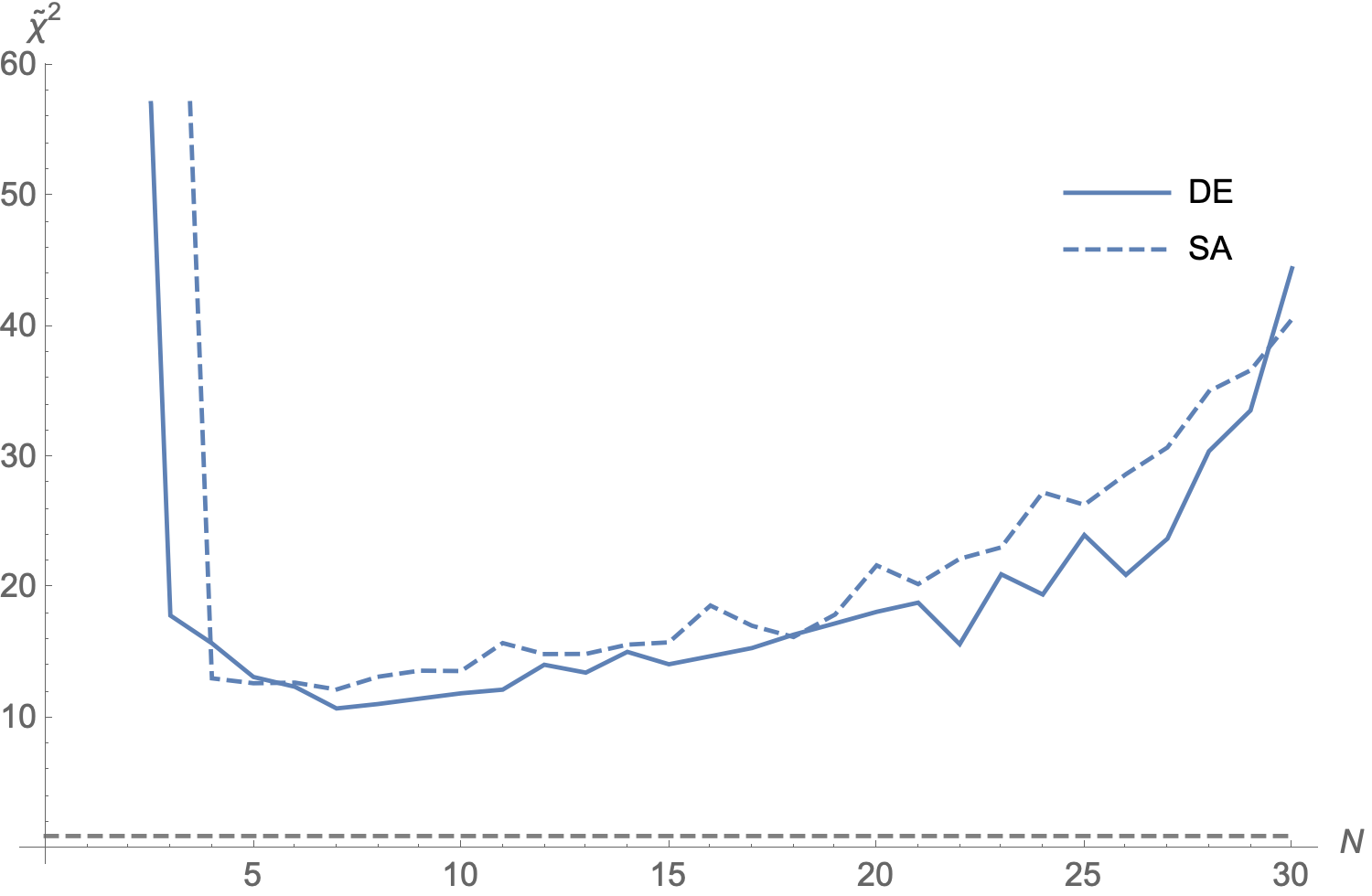}  
   \caption{Minimum of the $\chi^2/d.o.f.$ achieved by the global optimization methods for each of the lattice data sets. From top to bottom, the first line of plots refers to the $M_\pi = 150$ MeV,
   the second line to $M_\pi = 290$ MeV and the bottom line to $M_\pi = 422$ MeV. In each line, the left plot is the outcome of the minimization of the vector form factor, while the right plot refers
   to the minima for the scalar form factor.}
   \label{fig:quarkdata290}
\end{figure}

\begin{figure}[t]
   \centering
   \includegraphics[scale=0.38]{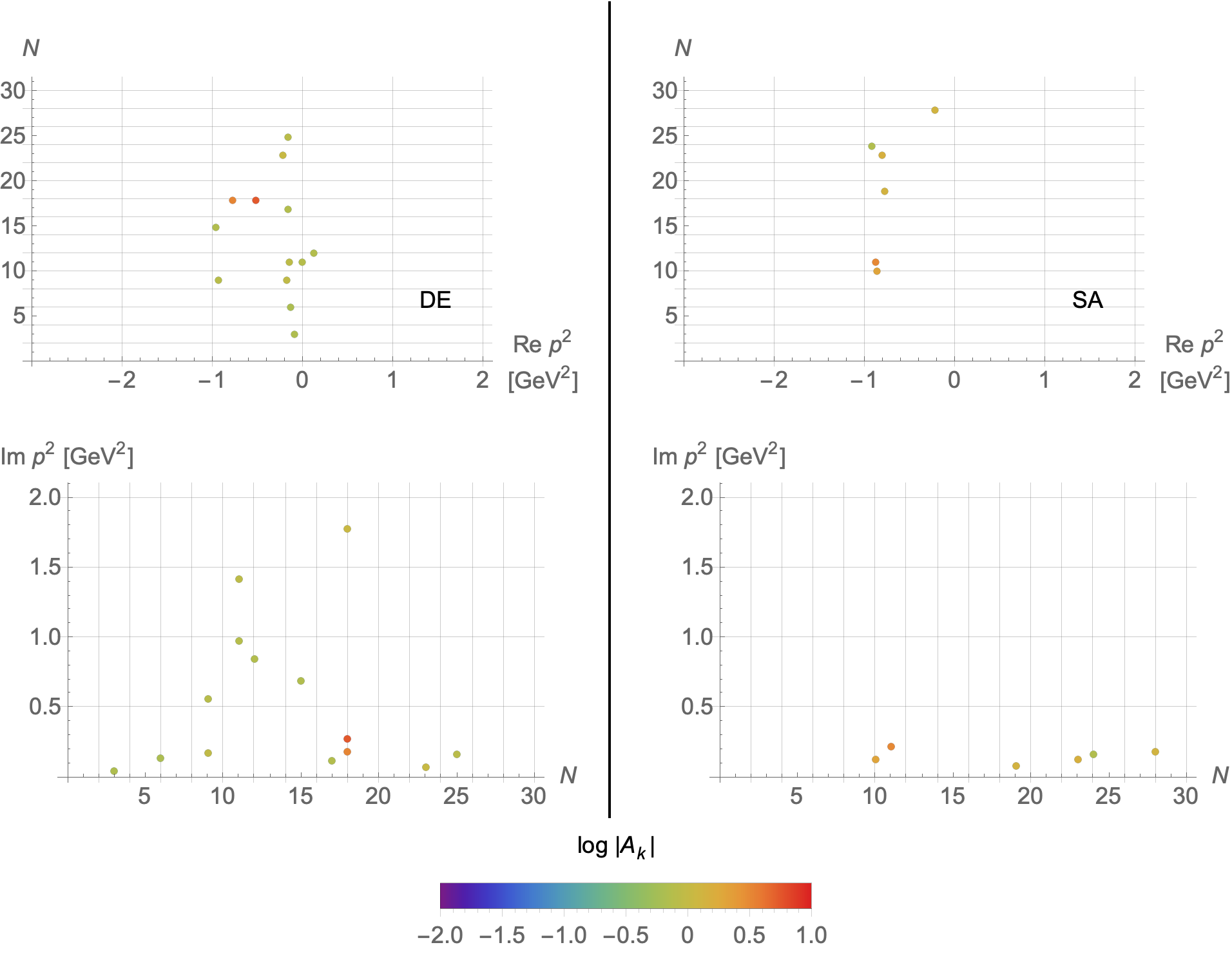} \\
   \includegraphics[scale=0.38]{OffaxisFv290.png}  
   \caption{Poles in absolute value of the residua for complex valued momenta for the simulation with a $M_\pi = 290$ MeV. On the upper plot the data refers to the
   vector form factor, while the lower plot refers to the analysis of the scalar form factor. Here the poles with $|Z| < 0.5$ are not shown.}
   \label{fig:quarkchi}
\end{figure}

\begin{figure}[t]
   \centering
   \includegraphics[scale=0.38]{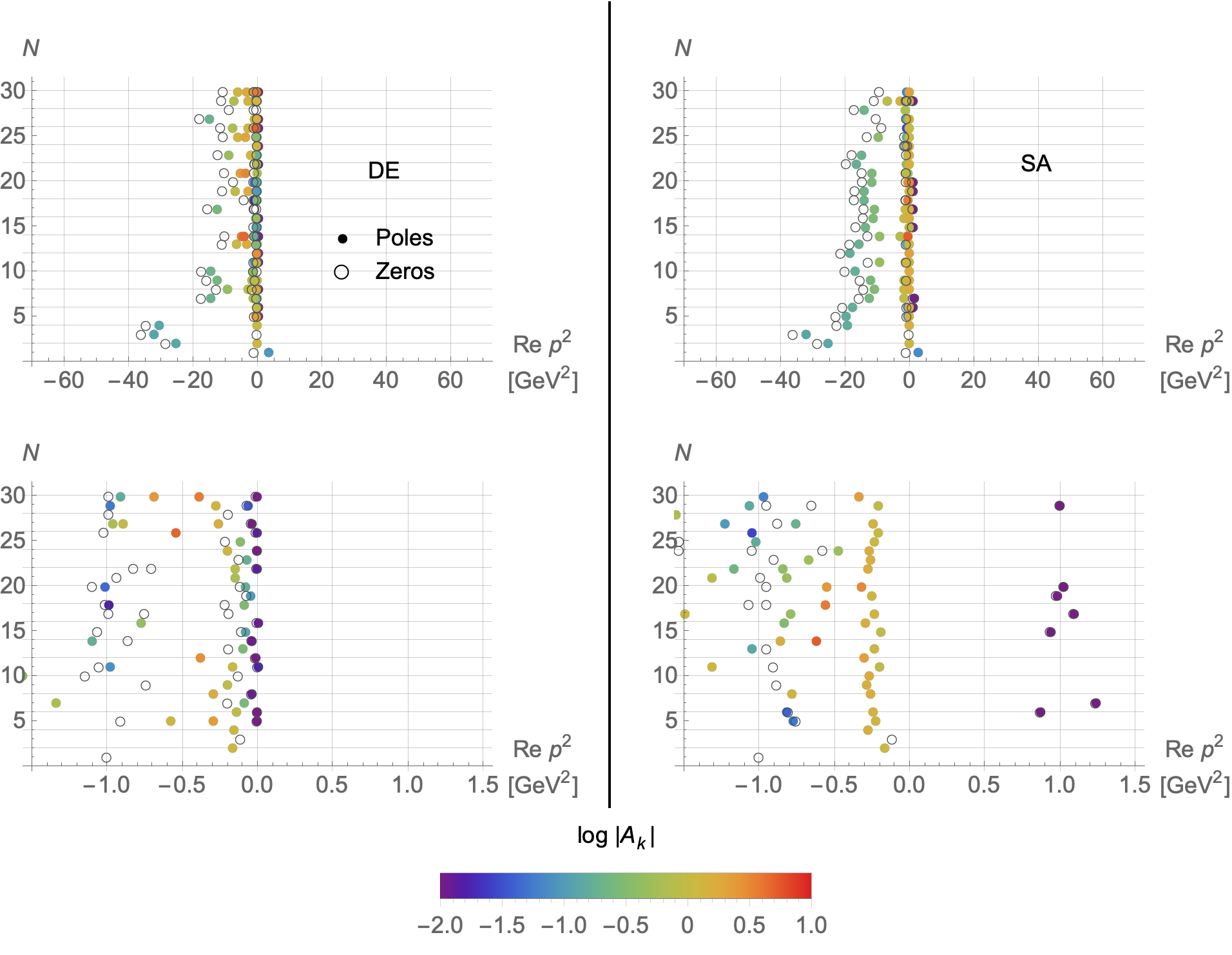} \\
   \includegraphics[scale=0.38]{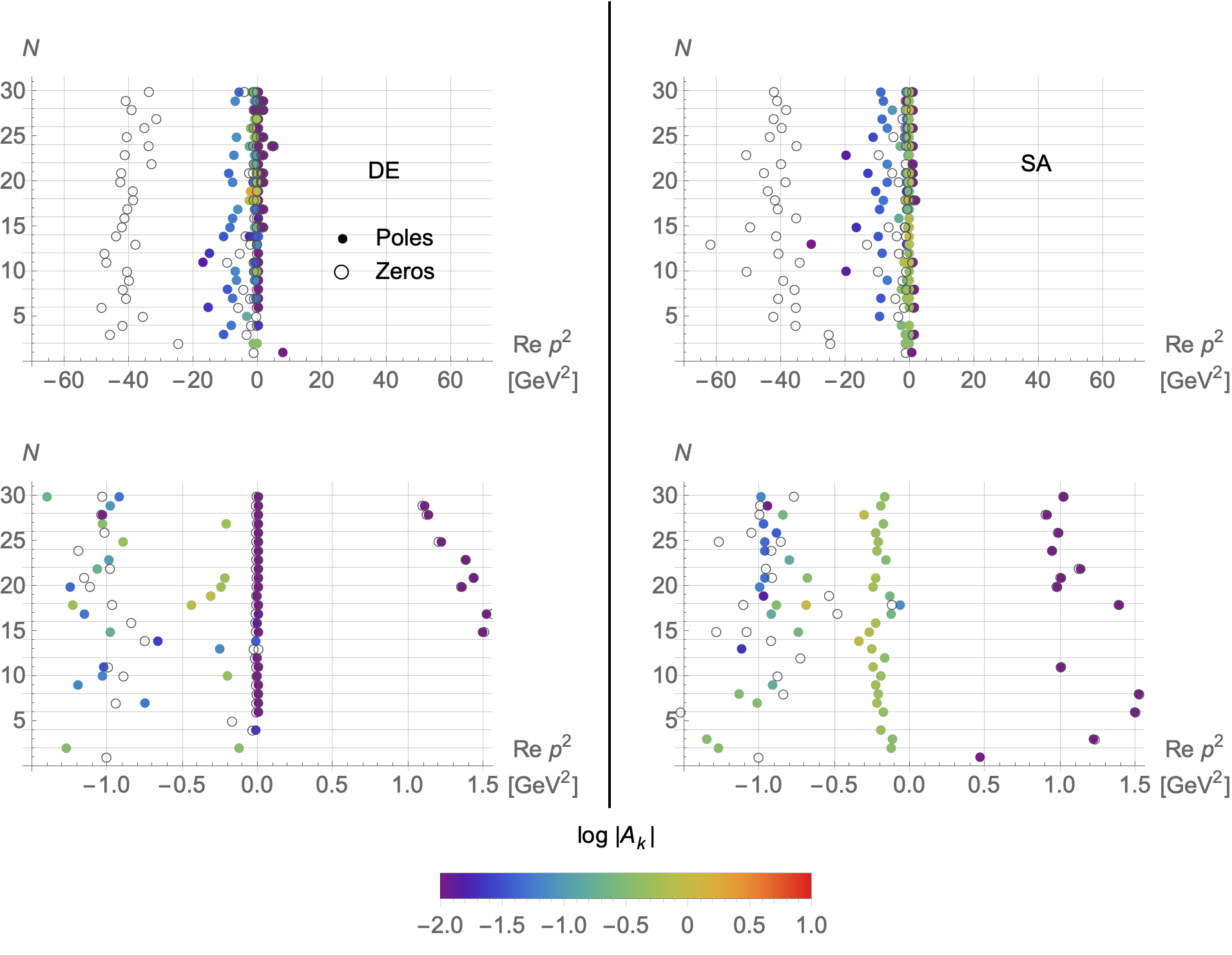}  
   \caption{Poles, zeros and residua  for real on-axis momenta for $M_\pi = 422$ MeV. The upper plot reports the results from the analysis of the
   vector form factor, while the lower plot reports the results of the analysis of the scalar part form factor. We call the reader attention that the same data appears twice for different regions of momenta. 
   This procedure is repeated in subsequent Figs.}
   \label{fig:quark422}
\end{figure}

The analysis of the Landau gauge lattice quark propagator assumes that this two point correlation function is color diagonal
and that its Lorentz-Dirac structure reads, in momentum space,
\begin{equation}
S(p) = Z(p^2) \frac{ \slashed{p} + M(p^2)}{p^2 + M^2(p^2)} = \frac{Z(p^2)}{D(p^2)} \, \Big( \slashed{p} + M(p^2) \Big)
\end{equation}
where $D(p^2) = p^2 + M^2(p^2)$. In the following, we name
\begin{equation}
 \frac{Z(p^2)}{D(p^2)} \qquad\mbox{ and }\qquad  \frac{ Z(p^2) \, M(p^2)}{D(p^2)}
\end{equation}
as vector and scalar form factors, respectively.
The lattice data for the propagator for the various ensembles can be seen in Fig, \ref{fig:quarkdata}. 
The vector and scalar form factors were rebuilt from the original lattice data assuming Gaussian distributions for the
propagation of errors.  
As see in Fig. \ref{fig:quarkdata}, the vector and scalar form factors are enhanced at low momenta and
the enhancement increases as the quark mass, or the pion mass, decreases. 
However, for the smaller quark mass, the scalar form factor drops faster as the momentum increases, with
the form factor becoming smaller than the two heavier pion masses considered for momenta $p \gtrsim 0.5$ GeV.
The comparison of the left and right plots in Fig. \ref{fig:quarkdata} reveals a complex situation that is the result of 
the dependence of the quark wave function $Z(p^2)$ and running quark mass $M(p^2)$ with the quark mass, i.e. with the pion mass.

The lattice data is to be described by a Pad\'e approximant where the coefficients of the polynomials are determined by minimizing the corresponding $\tilde{\chi}^2 = \chi^2/d.o.f.$ The  quality of the minimization 
can be measured by the value of $\tilde{\chi}^2 = \chi^2/d.o.f.$ at the minimum. These values, for the various sets of data, for the two
minimizating algorithms and for the various $N$  are reported in Fig. \ref{fig:quarkdata290}.
In general, the values of the $\tilde{\chi}^2$
obtained for the Pad\'e approximants are able to reproduce well the quark propagator form factors, with
the $\tilde{\chi}^2$
at the minimum associated with the simulated annealing method performing slightly worst than  for the differential evolution
method. The exception is the scalar data for the heaviest pion mass $M_\pi = 422$ MeV, whose $\tilde{\chi}^2 $
takes values well above the acceptable. 
In the following, we will disregard the data coming from the analysis of the scalar form factor associated with  $M_\pi = 422$ MeV.

The Pad\'e study of the poles, the zeros and the residua for complex momenta of the Pad\'e approximants shows no stable structures for momenta $|p^2| \leqslant 2$ Gev$^2$. 
The same was observed for the pure gauge ghost propagator in \cite{Falcao:2020vyr}.
This is illustrated in Fig. \ref{fig:quarkchi} for the simulation performed with $M_\pi = 290$ MeV. Similar plots can be show for the other two data sets. 
We take these results as an indication that the quark propagator has no poles for complex momenta. This configures a quite different nature for the analytic
structure of the fundamental QCD propagators. If the pure gauge gluon propagator has poles at complex momenta as suggest in \cite{Falcao:2020vyr,Binosi:2019ecz} 
and also by the good agreement found between the predictions based on the use of the Gribov-Zwanziger actions \cite{Zwanziger:1989mf,Dudal:2008sp,Dudal:2007cw}
and the lattice data for the infrared region
\cite{Dudal:2010tf,Oliveira:2012eh,Cucchieri:2011ig,Cucchieri:2016jwg,Dudal:2018cli}, the Pad\'e analysis suggests that the pure gauge ghost propagator and
the quark propagator are void of  poles for complex momenta. If this is the case, it remains to be understood the mechanism that in QCD distinguishes
the bosonic degrees of freedom from those whose nature is of the Grassmann type.

\begin{figure}[t]
   \centering
   \includegraphics[scale=0.38]{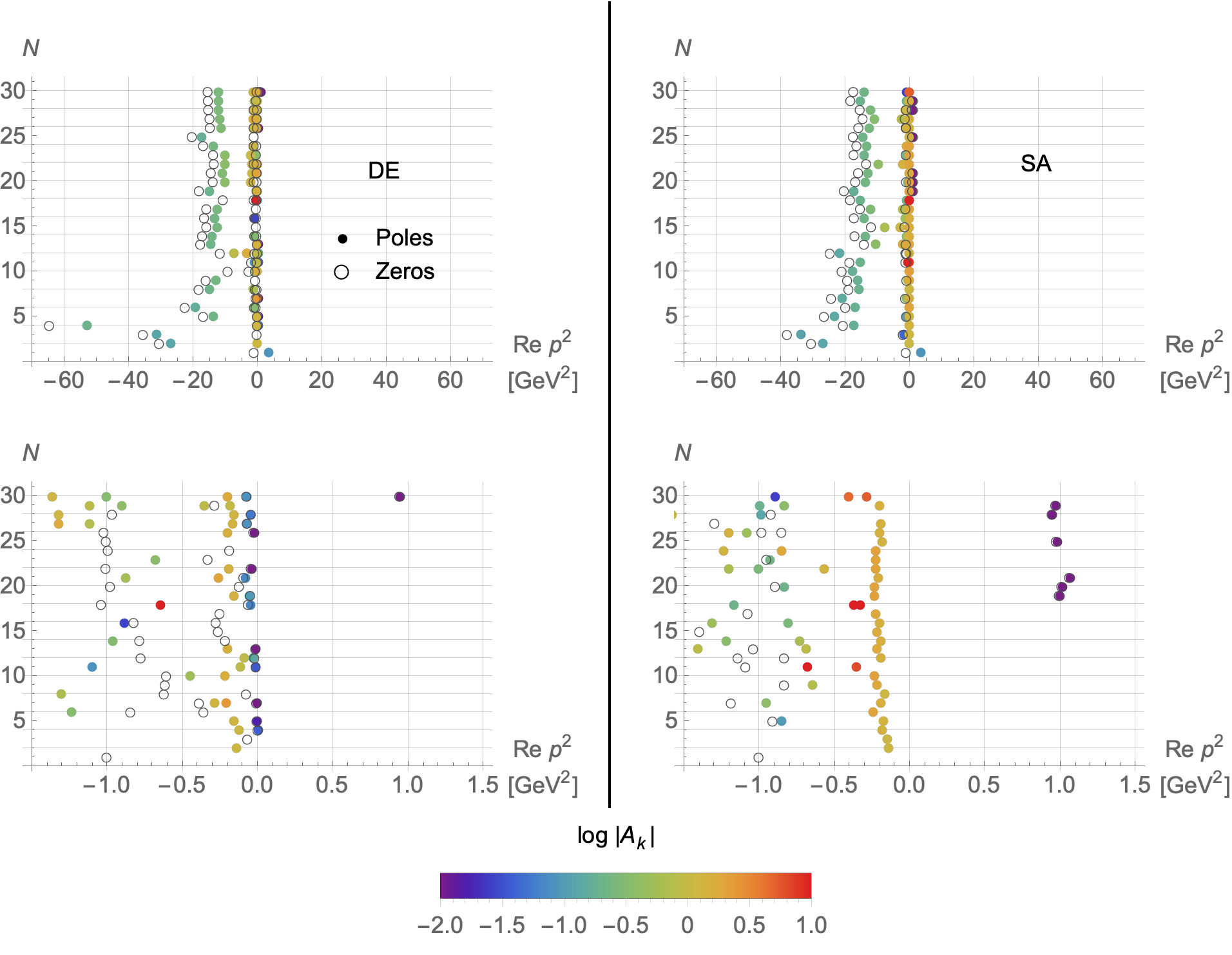} \\
   \includegraphics[scale=0.38]{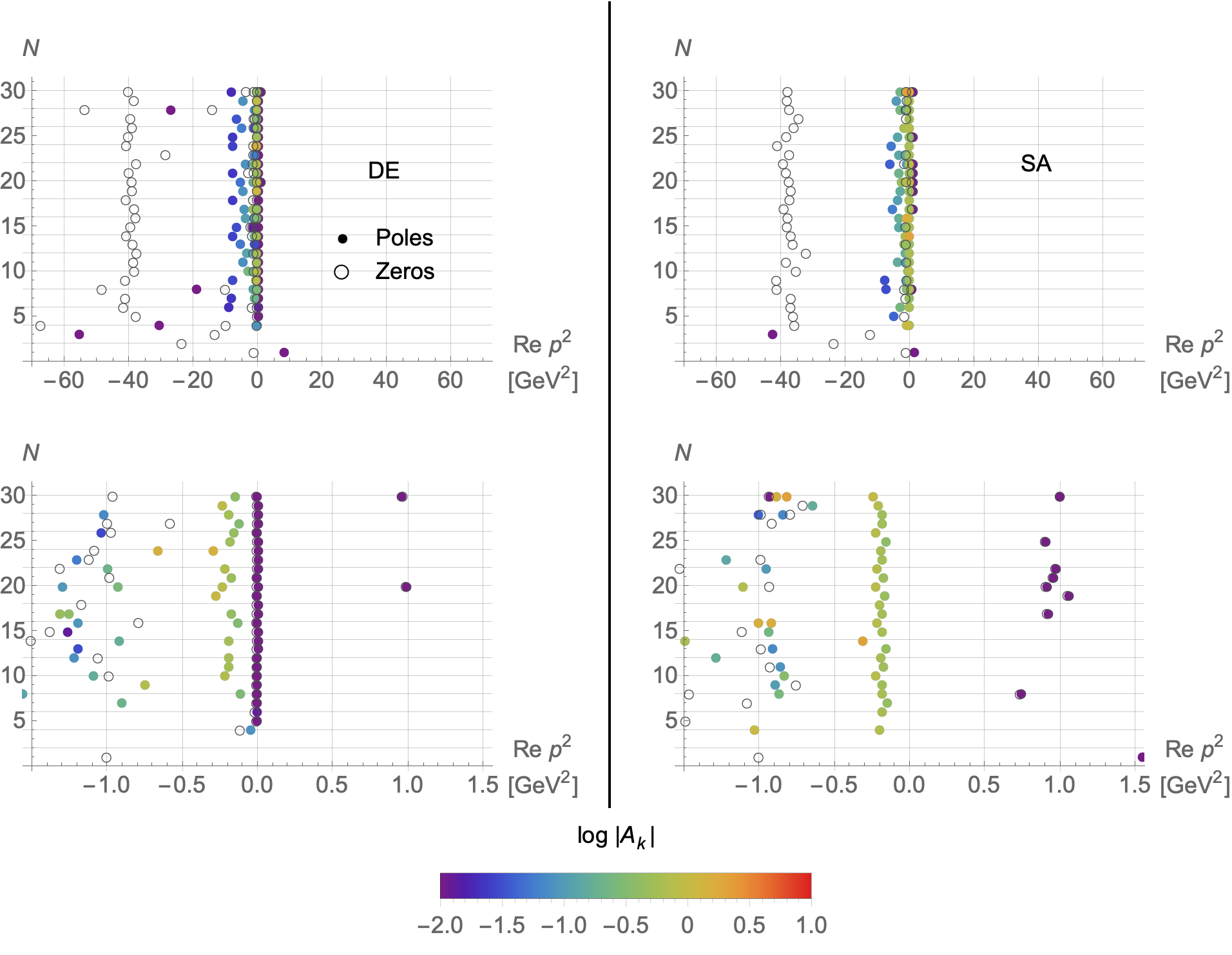}  
   \caption{The same as Fig. \ref{fig:quark422} but for the data associated with $M_\pi =290$ MeV.}
   \label{fig:quark290}
\end{figure}

\begin{figure}[t]
   \centering
   \includegraphics[scale=0.38]{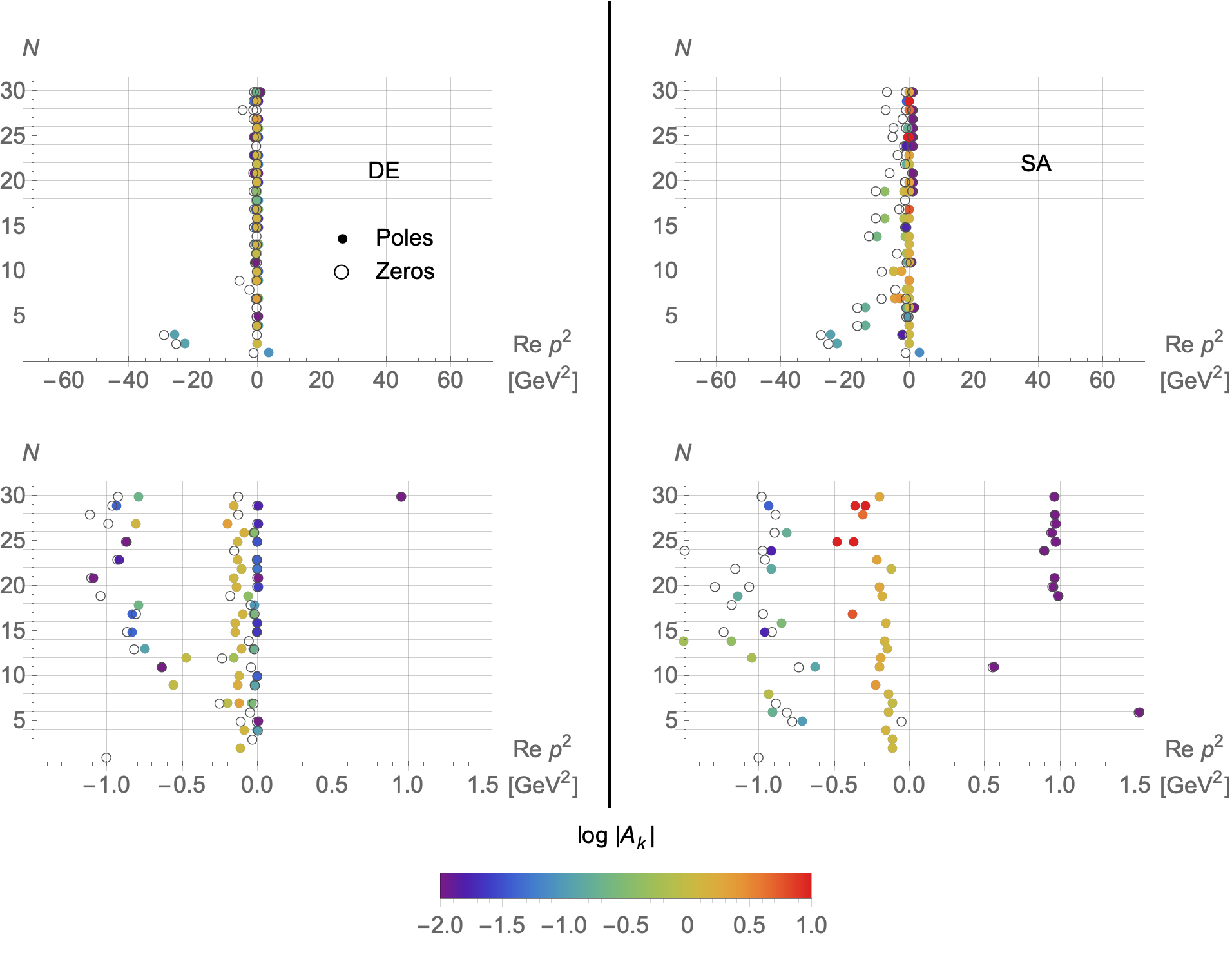} \\
   \includegraphics[scale=0.38]{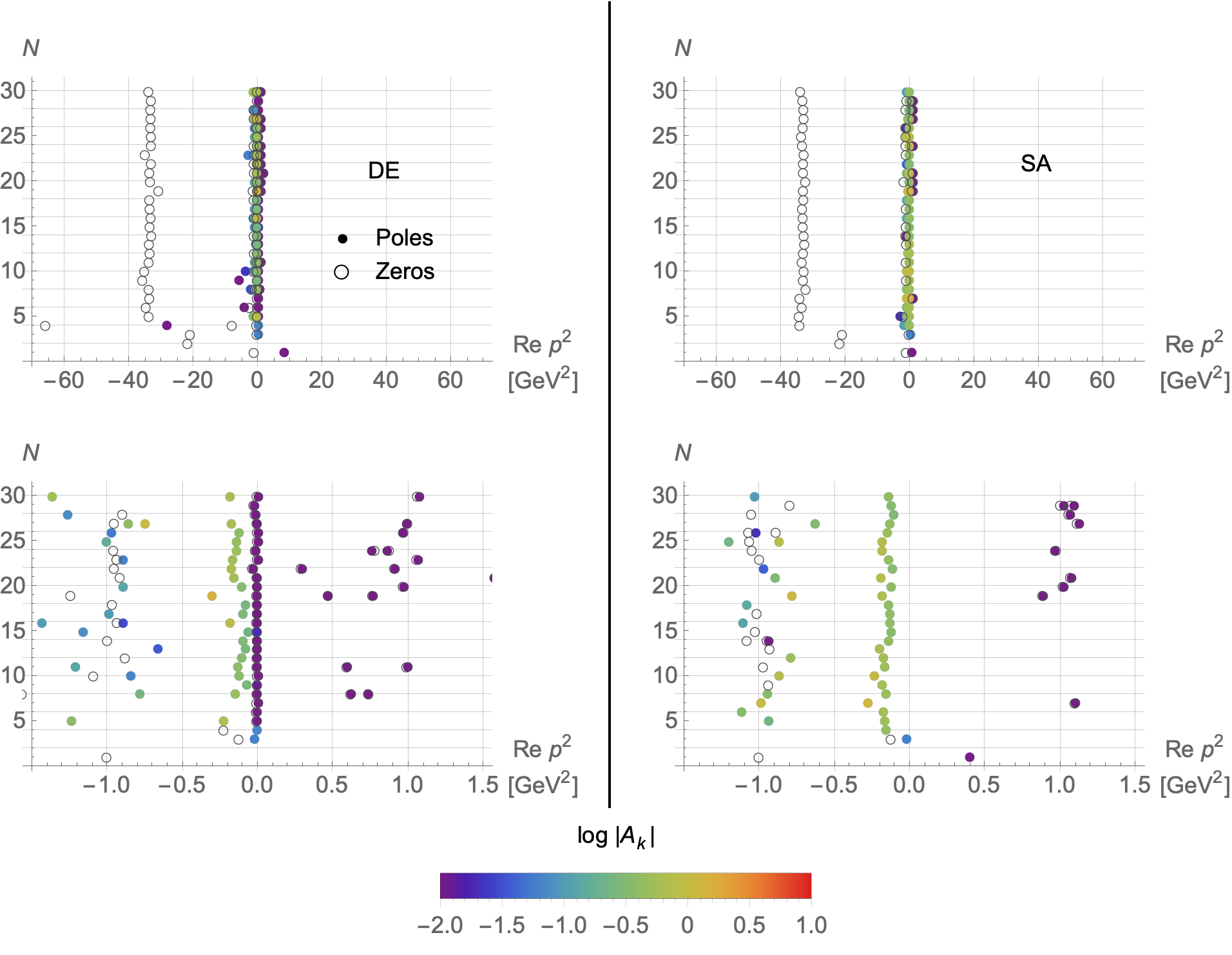}     
   \caption{The same as Figs \ref{fig:quark422} and \ref{fig:quark290} but for the data associated with $M_\pi =150$ MeV.}
   \label{fig:quark150}
\end{figure}

Let us turn now our attention to on-axis momenta. The outcome of the analysis of the form factors for $M_\pi = 422$ MeV can be seen in Fig.
\ref{fig:quark422}. In what concerns the vector form factor, the analysis of the results obtained with the differential evolution method does not have
 a clear interpretation. On the other hand, the results coming from the simulated annealing method reveal a pole at time-like
momenta $p^2 \sim -0.35$ GeV$^2$ whose position is essentially independent of the degree of Pad\'e approximant considered. The results for
the smaller $N$ where this pole appears, give a pole position at $p^2 = -0.275$ GeV$^2$ with a residuum $Z = 1.535$ when $N = 4$ and a pole at
$p^2 = -0.227$ GeV$^2$ with $Z = 1.266$ for $N =5$. The corresponding results obtained with the differential evolution method are a pole
$p^2 = -0.164$ GeV$^2$ and a residuum of $Z = 1.087$ for $N = 4$ and pole at $p^2 = -0.293$ GeV$^2$ with $Z = 2.325$ for $N= 5$ are
in the same ballpark as those obtained with the simulated annealing method. From the data it is difficult to identify any other structure as, for
example, a possible branch point. As already mentioned, the results coming from the analysis of the scalar form factor are not
so faithful. However, looking at the bottom plot of Fig. \ref{fig:quark422}, it qualitatively reproduces the results of the vector form factor
with the poles at slightly smaller values of $|p^2|$. This may suggest that the poles of the quark propagator do not
have to occur at the same momentum for its vectorial and scalar form factors. This  contrasts with the intuitive picture that
is built from the analysis of the free fermionic propagators as, for example, is found in QED or other theories where fermions  appear
as free particles.

The results of the analysis of the lattice data for the simulation with an $M_\pi = 290$ MeV are reported in Fig. \ref{fig:quark290}. 
For the vector form factor (upper plot), the outcome of the simulated annealing method is again easier to understand.
It shows a pole at time-like momenta, that is also observed with the differential evolution method. For the smallest values
of $N$, the simulated annealing returns a pole at $p^2 = -0.139$ GeV$^2$ and a residuum of $Z = 1.174$ for $N = 2$ and
$p^2 = -0.148$ GeV$^2$ with $Z = 1.201$ for $N = 3$, while the results of the differential evolution method
return a pole at $p^2 = -0.139$ GeV$^2$ with $Z = 1.174$ for $N = 2$ and
$p^2 = -0.127$ GeV$^2$ with $Z = 1.203$ for $N = 4$. No further clear poles or possible branch points are identified from the analysis
of the vector form factor. In what concerns the results for the scalar form factors, the two optimization methods give similar
results that sugest a branch point at $p^2 \sim -1$ GeV$^2$ where there is a proliferation of poles that have close zeros of the Pad\'e approximants.
A precise determination of a possible branch point would demand for less scattered data that, in principle, is possible to achieve with higher statistical
simulations. The data for the scalar form factor shows a clear pole whose location is at
$p^2 = -0.119$ GeV$^2$ with a residuum $Z = 0.299$ for $N = 8$ and
$p^2 = -0.223$ GeV$^2$ with $Z = 0.740$ for $N = 10$ according to the differential evolution method, and at
$p^2 = -0.205$ GeV$^2$ with a residuum $Z = 0.727$ for $N = 4$ and 
$p^2 = -0.182$ GeV$^2$ with $Z = 0.623$ for $N = 6$ by the simulated annealing method.
These results suggest that the pole of the vector form factor is at $p^2 = -0.2$ GeV$^2$ and its residuum $Z = 1.2$, while
the pole of the scalar form factor occurs at $p^2 = -0.2$ GeV and its residuum is $Z = 0.65$. Recall
that as for the largest pion mass, the pole residuum is positive defined and the pole of the vector and scalar form factors occur for the same time-like momentum.

Finally, the results of the analysis of the lattice data set with $M_\pi = 150$ MeV is reported in Fig. \ref{fig:quark150}. 
The study of the vector form factor suggest a branch point at time-like momenta but at $|p^2|$ that are smaller than the
branch point found for the $M_\pi = 290$ MeV data. However, once more, its precise location is difficult to determine within the statistical precision of
the simulation. On the other hand, both the differential evolution and the simulated annealing methods point towards the presence of a pole
at $p^2 = -0.114$ GeV$^2$ with $Z = 1.170$ for $N = 2$ or $p^2 = -0.094$ GeV$^2$ with $Z = 1.268$ for $N = 4$ (differential evolution), or
$p^2 = -0.114$ GeV$^2$ with $Z = 1.170$ for $N = 2$ and $p^2 = -0.117$ GeV$^2$ with $Z = 1.179$ for $N = 3$ (simulated annealing), respectively.
In what concerns the analysis of the scalar form factor, the results suggest the presence of branch point at similar values of time-like momenta as the found for the vector
form factor data. However, once more,  its precise location is hard to define. 
This pole occurs for $p^2 = -0.153$ GeV$^2$ with a $Z =  0.468$ for $N = 8$ and $p^2 = -0.122$ GeV$^2$ with $Z = 0.368$ for $N = 10$, according to the
differential evolution method, and at $p^2 = -0.161$ GeV$^2$ with $Z = 0.463$ for $N = 4$ and $p^2 = -0.170$ GeV$^2$ for $Z = 0.522$ for $N = 5$ when using the simulated annealing method.
Gathering these results together, one can claim a pole  at time-like momenta $p^2 = -0.11$ GeV$^2$ with a residuum $Z =1.17$ from  the vector form factor data,
and a pole at $p^2 = -0.15$ GeV$^2$ for $Z = 0.45$ from the scalar form factor data. Once more, the residuum of the quark propagator at the pole is
positive defined.

\begin{figure}[t]
   \centering
   \includegraphics[scale=0.38]{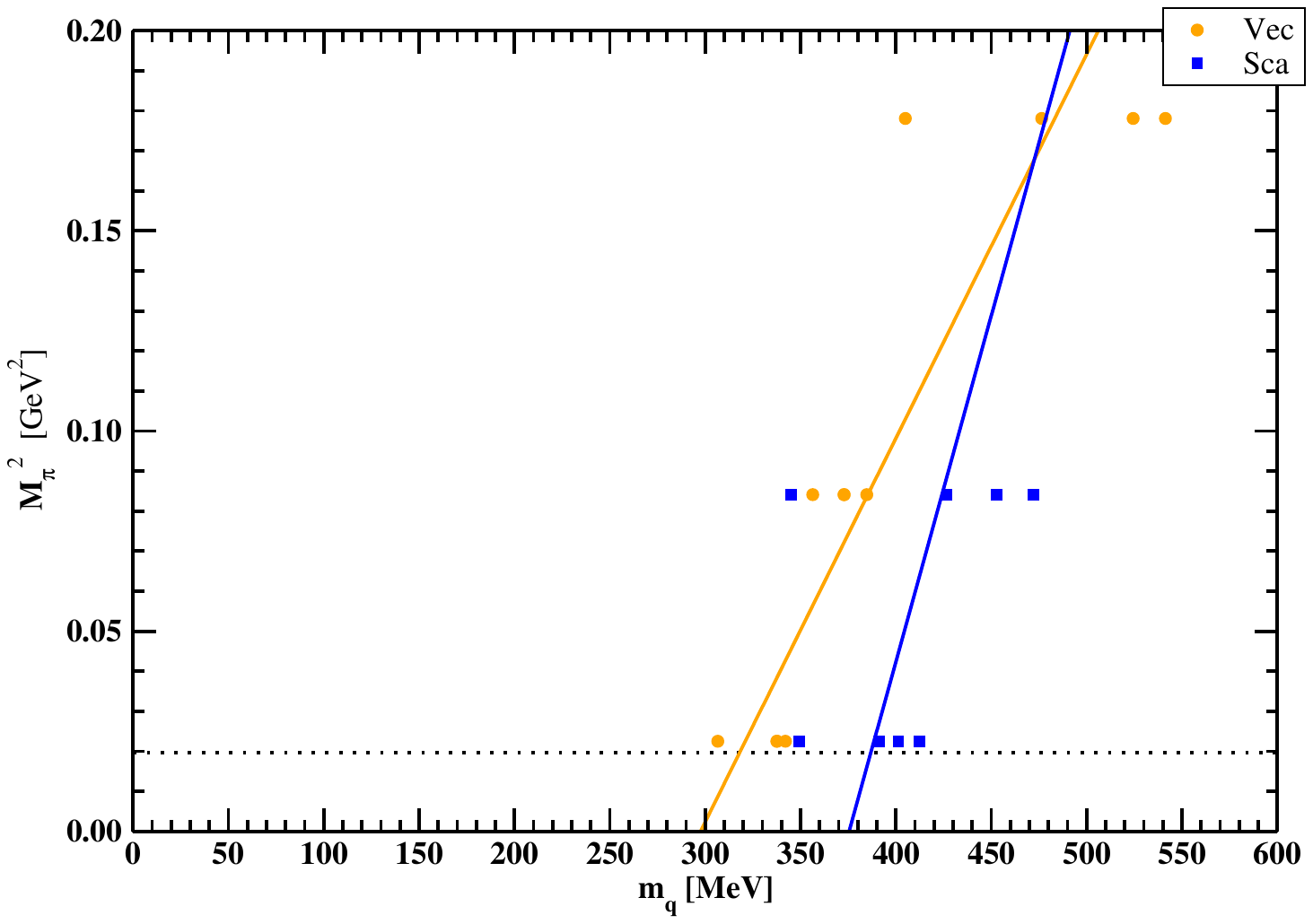} 
   \caption{Pion mass squared as a function of the pole quark mass as measured from vector and scalar form factors. The dotted line refers
   to the physical pion mass taken as $M_\pi = 140$ MeV.  
   To highlight the linear dependence of the pion mass squared on the quark mass we include the solid lines that are built using the average values of the 
   quark masses for each of the pion mass values and performing a linear regression (vector form factor in orange)  or going through the data points (scalar form factor in blue).
   Note that the errors on the estimation of the ``pole quark mass'' are quite large and, therefore, the continuum lines that are represented try only to illustrate the quark mass 
   dependence and appear in the Fig. to guide the eye of the reader.}
   \label{fig:MquarkMpion}
\end{figure}

The results just described suggest also that there is a correlation between the pion mass and the quark mass taken from the
dominant pole, i.e. the pole with the largest residuum, for real on-axis momenta. Defining the quark mass as 
$m^2_q = -p^2$, where $p^2$ is the value of pole position for real on-axis momenta, this correlation can be used to check
one of the most fundamental relations of QCD, namely the prediction based on the partial conservation of the axial current, see e.g. \cite{Yndurain:2006amm},
that gives
\begin{equation}
  M^2_\pi ~ \propto ~ m_q \ . \label{Eq:PCAC}
\end{equation}
The function $M^2_\pi ( m_q )$ is reported in Fig. \ref{fig:MquarkMpion} for the various estimates of $m_q$ mentioned previously and it also includes the physical value
of the pion mass taken to be 140 MeV. For the various $m_q$ associated with a given $M_\pi$ we take the pole mass for the smaller $N$ in the Pad\'e sequences.
There are two key points that can be read from Fig. \ref{fig:MquarkMpion}. The first being that, despite the large errors, the data reported is compatible with the predictions of PCAC
given in Eq. (\ref{Eq:PCAC}). The other point being that the pole mass for the scalar and vector form factors seems to be slightly different, with the pole position of the scalar
form part of the propagator appearing at slightly higher values of $m_q$. 

\section{Summary and Conclusions \label{Sec:summary}}

The investigation of the poles and branch cuts using Pad\'e approximants for the Landau gauge pure Yang-Mills gluon and ghost lattice propagators and for the full QCD quark propagator 
show significant differences between them. Indeed, only for the gluon case the method clearly identifies poles at complex momenta. 
The presence of complex poles associated with the gluon propagator were already seen in \cite{Dudal:2010tf}, where the compatibility of the predictions of Gribov-Zwanziger type of actions,
see \cite{Vandersickel:2012tz} and references therein, with the propagators calculated with lattice QCD simulations was investigated. In this sense, the observation of complex poles for the gluon 
propagator favors the Gribov-Zwanziger type of approach to the quantization of non-Abelian theories but it also rises questions about the definition of a proper quantum theory for non-Abelian theories; 
see e.g. the discussion in \cite{Hayashi:2021nnj,Hayashi:2021jju} and references therein. 
The absence of complex conjugate poles for the pure Yang-Millgs ghost and full QCD quark propagators requires  understanding the dynamics of the theory.
If only one of the fundamental QCD propagators has poles at complex momenta, it implies that a delicate tunning has to take place that prevents complex poles in the ghost and quark propagators.
Looking at the Dyson-Schwinger equations for the quark propagator and for the ghost propagtor, the mechanism responsible for the absence of the complex poles for these two propagator 
should translate in some type of constrains for the full ghost-gluon vertex and the full quark-gluon vertex. Recall that the aforementioned Dyson-Schwinger equations involve only these two vertices besides the propagators. 

The known solutions of the Dyson-Schwinger equations for the quark propagator suggest the presence of poles at complex momenta that the Pad\'e method is unable to identify. 
If this is a limitation of using Pad\'e approximants to look at the analytic structure of the propagators, or it is due to the statistical precision of the simulations,
or a limitation of the available analysis of the Dyson-Schwinger gap equation remains to be investigated. 
It is well known that the analytical structure predicted by the Dyson-Schwinger equations for the the quark propagator dependes, to some extend, on how the quark-gluon vertex is dressed.
Despite this difference, the quark and the ghost propagators computed from these equations are in good agreement with the outcome of the lattice simulations, suggesting that the 
description of the fundamental QCD propagators with continuum functional methods captures the essential of the dynamics of the theory.

The Pad\'e analysis of the Landau gauge quark propagator identifies also a pole for Minkowski momentum, i.e. real on-axis negative Euclidean momentum,
with a positive residuum. The naive interpretation of such a pole translates into a free asymptotic single quark state. However, it is well known that positive violation occurs for the quark propagator,
i.e. the spectral density of the quark is not positive defined, and the quark propagator can only be described by a single pole in addition with other structures that, overall, prevent the quarks
to appear as free particles. Unfortunately, the method used herein does not provide information on the structure beyond the single pole of the fermionic propagator. 

The quark propagator pole position in momentum space  can be translated into an effective quark mass.
This effective mass falls is in the range of values that, typically, are associated with the effective quark mass used in quark models. As can be seen in Fig. \ref{fig:MquarkMpion},
this effective quark mass is above 300 MeV. Moreover, the same Fig. suggests that the pole position can differ for the scalar and vector form factors, with the
scalar form factors favouring slightly large values for the effective mass. This result is somehow unexpected and maybe an artefact of the method. Indeed, assuming that there is a spectral
representation for the quark propagator, then it can be written as (ignoring the color part)
\begin{equation}
   S(p) = \int_{0}^{+ \infty} ~d \mu ~ \frac{ \slashed{p} \, \rho_1 (\mu) + m  \, \rho_2 ( \mu ) }{ p^2 - \mu + i \,  \epsilon} \ ,
\end{equation}
where $\rho_1 ( \mu )$ and $\rho_2 ( \mu )$ are the quark spectral functions \cite{Delbourgo:1977jc}. 
From the above expression it follows that
to accommodate a different pole structure for the vector and scalar parts of the propagator
the functions $\rho_1 ( \mu )$ and $\rho_2 ( \mu )$ have to be different.
Alternatively, this could be an indication of the absence of such  a type of integral representation for the propagator. 
One should also take into consideration the large uncertainties on the estimation of
the effective mass and that it is not clear on how to estimate the errors for the effective mass.

The results of the Pad\'e analysis also show a correlation between the effective quark mass and the corresponding pion mass. Furthermore, it is shown in Fig. \ref{fig:MquarkMpion} that this correlation,
measured by the curve $M^2_\pi (m_q)$,  is compatible with the predictions of partial conservation of the axial current for QCD.

The work described in the current manuscript and in \cite{Falcao:2020vyr} for the pure gauge theory propagators shows that one can rely on Pad\'e approximants combined with global optimisation techniques
to access the analytic structure of the two point correlation functions. The accuracy on the results depends on the size of the statistical ensemble of configurations and, certainly, having access to the simulations with
larger sets of gauge configurations will clear the outcome of the analysis.

\acknowledgements

The authors  where supported by national funds from FCT – Funda\c{c}\~ao para a  Ci\^encia e aTecnologia, I. P., within the projects UIDB/04564/2020, UIDP/04564/2020.
A.F.F. acknowledges the financial support via the Starting Grant from Trond Mohn Foundation (BFS2018REK01) and the University of Bergen.



\begin{thebibliography}{99}

\bibitem{Alkofer:2000wg}
R.~Alkofer and L.~von Smekal,
Phys. Rept. \textbf{353}, 281 (2001)
doi:10.1016/S0370-1573(01)00010-2
[arXiv:hep-ph/0007355 [hep-ph]].

\bibitem{Fischer:2006ub}
C.~S.~Fischer,
J. Phys. G \textbf{32}, R253-R291 (2006)
doi:10.1088/0954-3899/32/8/R02
[arXiv:hep-ph/0605173 [hep-ph]].

\bibitem{Binosi:2009qm}
D.~Binosi and J.~Papavassiliou,
Phys. Rept. \textbf{479}, 1-152 (2009)
doi:10.1016/j.physrep.2009.05.001
[arXiv:0909.2536 [hep-ph]].


\bibitem{Perl:2009zz}
M.~L.~Perl, E.~R.~Lee and D.~Loomba,
Ann. Rev. Nucl. Part. Sci. \textbf{59}, 47-65 (2009)
doi:10.1146/annurev-nucl-121908-122035

\bibitem{Tanabashi:2018oca}
M.~Tanabashi \textit{et al.} [Particle Data Group],
Phys. Rev. D \textbf{98}, no.3, 030001 (2018)
doi:10.1103/PhysRevD.98.030001


\bibitem{Papavassiliou:2022wrb}
J.~Papavassiliou,
Chin. Phys. C \textbf{46}, no.11, 112001 (2022)
doi:10.1088/1674-1137/ac84ca
[arXiv:2207.04977 [hep-ph]].

\bibitem{Leinweber:1998uu}
D.~B.~Leinweber \textit{et al.} [UKQCD],
Phys. Rev. D \textbf{60}, 094507 (1999)
[erratum: Phys. Rev. D \textbf{61}, 079901 (2000)]
doi:10.1103/PhysRevD.60.094507
[arXiv:hep-lat/9811027 [hep-lat]].

\bibitem{Becirevic:1999uc}
D.~Becirevic, P.~Boucaud, J.~P.~Leroy, J.~Micheli, O.~Pene, J.~Rodriguez-Quintero and C.~Roiesnel,
Phys. Rev. D \textbf{60}, 094509 (1999)
doi:10.1103/PhysRevD.60.094509
[arXiv:hep-ph/9903364 [hep-ph]].

\bibitem{Becirevic:1999hj}
D.~Becirevic, P.~Boucaud, J.~P.~Leroy, J.~Micheli, O.~Pene, J.~Rodriguez-Quintero and C.~Roiesnel,
Phys. Rev. D \textbf{61}, 114508 (2000)
doi:10.1103/PhysRevD.61.114508
[arXiv:hep-ph/9910204 [hep-ph]].

\bibitem{Cucchieri:2007md}
A.~Cucchieri and T.~Mendes,
PoS \textbf{LATTICE2007}, 297 (2007)
doi:10.22323/1.042.0297
[arXiv:0710.0412 [hep-lat]].

\bibitem{Bogolubsky:2009dc}
I.~L.~Bogolubsky, E.~M.~Ilgenfritz, M.~Muller-Preussker and A.~Sternbeck,
Phys. Lett. B \textbf{676}, 69-73 (2009)
doi:10.1016/j.physletb.2009.04.076
[arXiv:0901.0736 [hep-lat]].

\bibitem{Dudal:2010tf}
D.~Dudal, O.~Oliveira and N.~Vandersickel,
Phys. Rev. D \textbf{81}, 074505 (2010)
doi:10.1103/PhysRevD.81.074505
[arXiv:1002.2374 [hep-lat]].

\bibitem{Cucchieri:2011ig}
A.~Cucchieri, D.~Dudal, T.~Mendes and N.~Vandersickel,
Phys. Rev. D \textbf{85}, 094513 (2012)
doi:10.1103/PhysRevD.85.094513
[arXiv:1111.2327 [hep-lat]].

\bibitem{Maas:2011ez}
A.~Maas, J.~M.~Pawlowski, L.~von Smekal and D.~Spielmann,
Phys. Rev. D \textbf{85}, 034037 (2012)
doi:10.1103/PhysRevD.85.034037
[arXiv:1110.6340 [hep-lat]].

\bibitem{Dudal:2018cli}
D.~Dudal, O.~Oliveira and P.~J.~Silva,
Annals Phys. \textbf{397}, 351-364 (2018)
doi:10.1016/j.aop.2018.08.019
[arXiv:1803.02281 [hep-lat]].

\bibitem{Li:2019hyv}
S.~W.~Li, P.~Lowdon, O.~Oliveira and P.~J.~Silva,
Phys. Lett. B \textbf{803}, 135329 (2020)
doi:10.1016/j.physletb.2020.135329
[arXiv:1907.10073 [hep-th]].

\bibitem{Catumba:2021hcx}
G.~T.~R.~Catumba, O.~Oliveira and P.~J.~Silva,
Phys. Rev. D \textbf{103}, no.7, 074501 (2021)
doi:10.1103/PhysRevD.103.074501
[arXiv:2101.04978 [hep-lat]].


\bibitem{Boucaud:2007hy}
P.~Boucaud, J.~P.~Leroy, A.~Le Yaouanc, A.~Y.~Lokhov, J.~Micheli, O.~Pene, J.~Rodriguez-Quintero and C.~Roiesnel,
JHEP \textbf{03}, 076 (2007)
doi:10.1088/1126-6708/2007/03/076
[arXiv:hep-ph/0702092 [hep-ph]].

\bibitem{Huber:2007kc}
M.~Q.~Huber, R.~Alkofer, C.~S.~Fischer and K.~Schwenzer,
Phys. Lett. B \textbf{659}, 434-440 (2008)
doi:10.1016/j.physletb.2007.10.073
[arXiv:0705.3809 [hep-ph]].

\bibitem{Dudal:2008rm}
D.~Dudal, J.~A.~Gracey, S.~P.~Sorella, N.~Vandersickel and H.~Verschelde,
Phys. Rev. D \textbf{78}, 125012 (2008)
doi:10.1103/PhysRevD.78.125012
[arXiv:0808.0893 [hep-th]].

\bibitem{Aguilar:2008xm}
A.~C.~Aguilar, D.~Binosi and J.~Papavassiliou,
Phys. Rev. D \textbf{78}, 025010 (2008)
doi:10.1103/PhysRevD.78.025010
[arXiv:0802.1870 [hep-ph]].

\bibitem{Rodriguez-Quintero:2010qad}
J.~Rodriguez-Quintero,
JHEP \textbf{01}, 105 (2011)
doi:10.1007/JHEP01(2011)105
[arXiv:1005.4598 [hep-ph]].

\bibitem{Aguilar:2010zx}
A.~C.~Aguilar, D.~Binosi and J.~Papavassiliou,
Phys. Rev. D \textbf{81}, 125025 (2010)
doi:10.1103/PhysRevD.81.125025
[arXiv:1004.2011 [hep-ph]].

\bibitem{Strauss:2012zz}
S.~Strauss, C.~S.~Fischer and C.~Kellermann,
Prog. Part. Nucl. Phys. \textbf{67}, 239-244 (2012)
doi:10.1016/j.ppnp.2011.12.025

\bibitem{Pelaez:2014mxa}
M.~Pel\'aez, M.~Tissier and N.~Wschebor,
Phys. Rev. D \textbf{90}, 065031 (2014)
doi:10.1103/PhysRevD.90.065031
[arXiv:1407.2005 [hep-th]].

\bibitem{Aguilar:2019uob}
A.~C.~Aguilar, F.~De Soto, M.~N.~Ferreira, J.~Papavassiliou, J.~Rodr\'\i{}guez-Quintero and S.~Zafeiropoulos,
Eur. Phys. J. C \textbf{80}, no.2, 154 (2020)
doi:10.1140/epjc/s10052-020-7741-0
[arXiv:1912.12086 [hep-ph]].

\bibitem{Reinosa:2020skx}
U.~Reinosa, J.~Serreau, R.~C.~Terin and M.~Tissier,
SciPost Phys. \textbf{10}, no.2, 035 (2021)
doi:10.21468/SciPostPhys.10.2.035
[arXiv:2004.12413 [hep-th]].

\bibitem{Fischer:2020xnb}
C.~S.~Fischer and M.~Q.~Huber,
Phys. Rev. D \textbf{102}, no.9, 094005 (2020)
doi:10.1103/PhysRevD.102.094005
[arXiv:2007.11505 [hep-ph]].

\bibitem{Eichmann:2021zuv}
G.~Eichmann, J.~M.~Pawlowski and J.~M.~Silva,
Phys. Rev. D \textbf{104}, no.11, 114016 (2021)
doi:10.1103/PhysRevD.104.114016
[arXiv:2107.05352 [hep-ph]].

\bibitem{Pelaez:2021tpq}
M.~Pel\'aez, U.~Reinosa, J.~Serreau, M.~Tissier and N.~Wschebor,
Rept. Prog. Phys. \textbf{84}, no.12, 124202 (2021)
doi:10.1088/1361-6633/ac36b8
[arXiv:2106.04526 [hep-th]].

\bibitem{Dudal:2022nnu}
D.~Dudal, D.~M.~van Egmond, U.~Reinosa and D.~Vercauteren,
[arXiv:2206.06002 [hep-th]].

\bibitem{vonSmekal:1997ohs}
L.~von Smekal, R.~Alkofer and A.~Hauck,
Phys. Rev. Lett. \textbf{79}, 3591-3594 (1997)
doi:10.1103/PhysRevLett.79.3591
[arXiv:hep-ph/9705242 [hep-ph]].


\bibitem{Alkofer:2003jj}
R.~Alkofer, W.~Detmold, C.~S.~Fischer and P.~Maris,
Phys. Rev. D \textbf{70}, 014014 (2004)
doi:10.1103/PhysRevD.70.014014
[arXiv:hep-ph/0309077 [hep-ph]].


\bibitem{Morris:2005tv}
T.~R.~Morris and O.~J.~Rosten,
Phys. Rev. D \textbf{73}, 065003 (2006)
doi:10.1103/PhysRevD.73.065003
[arXiv:hep-th/0508026 [hep-th]].

\bibitem{Morris:2006in}
T.~R.~Morris and O.~J.~Rosten,
J. Phys. A \textbf{39}, 11657-11681 (2006)
doi:10.1088/0305-4470/39/37/020
[arXiv:hep-th/0606189 [hep-th]].

\bibitem{Boucaud:2007va}
P.~Boucaud, J.~P.~Leroy, A.~Le Yaouanc, A.~Y.~Lokhov, J.~Micheli, O.~Pene, J.~Rodriguez-Quintero and C.~Roiesnel,
Eur. Phys. J. A \textbf{31}, 750-753 (2007)
doi:10.1140/epja/i2006-10295-1
[arXiv:hep-ph/0701114 [hep-ph]].


\bibitem{Arnone:2005fb}
S.~Arnone, T.~R.~Morris and O.~J.~Rosten,
Eur. Phys. J. C \textbf{50}, 467-504 (2007)
doi:10.1140/epjc/s10052-007-0258-y
[arXiv:hep-th/0507154 [hep-th]].

\bibitem{Zwanziger:2012xg}
D.~Zwanziger,
Phys. Rev. D \textbf{87}, 085039 (2013)
doi:10.1103/PhysRevD.87.085039
[arXiv:1209.1974 [hep-ph]].

\bibitem{Allendes:2014fua}
P.~Allendes, C.~Ayala and G.~Cveti\v{c},
Phys. Rev. D \textbf{89}, no.5, 054016 (2014)
doi:10.1103/PhysRevD.89.054016
[arXiv:1401.1192 [hep-ph]].

\bibitem{Meyers:2014iwa}
J.~Meyers and E.~S.~Swanson,
Phys. Rev. D \textbf{90}, no.4, 045037 (2014)
doi:10.1103/PhysRevD.90.045037
[arXiv:1403.4350 [hep-ph]].

\bibitem{Siringo:2014lva}
F.~Siringo,
Phys. Rev. D \textbf{90}, no.9, 094021 (2014)
doi:10.1103/PhysRevD.90.094021
[arXiv:1408.5313 [hep-ph]].

\bibitem{Dudal:2015khv}
D.~Dudal and M.~S.~Guimaraes,
Phys. Rev. D \textbf{93}, no.8, 085010 (2016)
doi:10.1103/PhysRevD.93.085010
[arXiv:1511.00902 [hep-th]].

\bibitem{Cyrol:2016tym}
A.~K.~Cyrol, L.~Fister, M.~Mitter, J.~M.~Pawlowski and N.~Strodthoff,
Phys. Rev. D \textbf{94}, no.5, 054005 (2016)
doi:10.1103/PhysRevD.94.054005
[arXiv:1605.01856 [hep-ph]].

\bibitem{Gogokhia:2016rix}
V.~Gogokhia and G.~G.~Barnaf\"oldi,
Int. J. Mod. Phys. A \textbf{31}, no.28\&29, 1645027 (2016)
doi:10.1142/S0217751X16450275

\bibitem{Cyrol:2017ewj}
A.~K.~Cyrol, M.~Mitter, J.~M.~Pawlowski and N.~Strodthoff,
Phys. Rev. D \textbf{97}, no.5, 054006 (2018)
doi:10.1103/PhysRevD.97.054006
[arXiv:1706.06326 [hep-ph]].


\bibitem{Cyrol:2018xeq}
A.~K.~Cyrol, J.~M.~Pawlowski, A.~Rothkopf and N.~Wink,
SciPost Phys. \textbf{5}, no.6, 065 (2018)
doi:10.21468/SciPostPhys.5.6.065
[arXiv:1804.00945 [hep-ph]].

\bibitem{Lowdon:2018mbn}
P.~Lowdon,
Phys. Lett. B \textbf{786}, 399-402 (2018)
doi:10.1016/j.physletb.2018.10.023
[arXiv:1801.09337 [hep-th]].

\bibitem{Siringo:2018uho}
F.~Siringo and G.~Comitini,
Phys. Rev. D \textbf{98}, no.3, 034023 (2018)
doi:10.1103/PhysRevD.98.034023
[arXiv:1806.08397 [hep-ph]].

\bibitem{Mintz:2018hhx}
B.~W.~Mintz, L.~F.~Palhares, G.~Peruzzo and S.~P.~Sorella,
Phys. Rev. D \textbf{99}, no.3, 034002 (2019)
doi:10.1103/PhysRevD.99.034002
[arXiv:1812.03166 [hep-th]].

\bibitem{Aguilar:2020uqw}
A.~C.~Aguilar, M.~N.~Ferreira and J.~Papavassiliou,
Eur. Phys. J. C \textbf{81}, no.1, 54 (2021)
doi:10.1140/epjc/s10052-021-08849-8
[arXiv:2010.12714 [hep-ph]].


\bibitem{Huber:2020keu}
M.~Q.~Huber,
Phys. Rev. D \textbf{101}, 114009 (2020)
doi:10.1103/PhysRevD.101.114009
[arXiv:2003.13703 [hep-ph]].


\bibitem{Napetschnig:2021ria}
M.~Napetschnig, R.~Alkofer, M.~Q.~Huber and J.~M.~Pawlowski,
Phys. Rev. D \textbf{104}, no.5, 054003 (2021)
doi:10.1103/PhysRevD.104.054003
[arXiv:2106.12559 [hep-ph]].


\bibitem{Horak:2021syv}
J.~Horak, J.~M.~Pawlowski, J.~Rodr\'\i{}guez-Quintero, J.~Turnwald, J.~M.~Urban, N.~Wink and S.~Zafeiropoulos,
Phys. Rev. D \textbf{105}, no.3, 036014 (2022)
doi:10.1103/PhysRevD.105.036014
[arXiv:2107.13464 [hep-ph]].


\bibitem{Aguilar:2021uwa}
A.~C.~Aguilar, M.~N.~Ferreira and J.~Papavassiliou,
Phys. Rev. D \textbf{105}, no.1, 014030 (2022)
doi:10.1103/PhysRevD.105.014030
[arXiv:2111.09431 [hep-ph]].


\bibitem{Gracey:2022xsz}
J.~A.~Gracey,
Phys. Rev. D \textbf{106}, no.6, 065006 (2022)
doi:10.1103/PhysRevD.106.065006
[arXiv:2207.04940 [hep-th]].

\bibitem{Horak:2022aqx}
J.~Horak, F.~Ihssen, J.~Papavassiliou, J.~M.~Pawlowski, A.~Weber and C.~Wetterich,
SciPost Phys. \textbf{13}, no.2, 042 (2022)
doi:10.21468/SciPostPhys.13.2.042
[arXiv:2201.09747 [hep-ph]].


\bibitem{Horak:2022myj}
J.~Horak, J.~M.~Pawlowski and N.~Wink,
[arXiv:2202.09333 [hep-th]].


\bibitem{Aguilar:2022wsh}
A.~C.~Aguilar, M.~N.~Ferreira, B.~M.~Oliveira and J.~Papavassiliou,
[arXiv:2210.07429 [hep-ph]].




\bibitem{Alkofer:2003jr}
R.~Alkofer, C.~S.~Fischer, H.~Reinhardt and L.~von Smekal,
Phys. Rev. D \textbf{68}, 045003 (2003)
doi:10.1103/PhysRevD.68.045003
[arXiv:hep-th/0304134 [hep-th]].

\bibitem{Duarte:2016iko}
A.~G.~Duarte, O.~Oliveira and P.~J.~Silva,
Phys. Rev. D \textbf{94}, no.1, 014502 (2016)
doi:10.1103/PhysRevD.94.014502
[arXiv:1605.00594 [hep-lat]].

\bibitem{Boucaud:2017ksi}
P.~Boucaud, F.~De Soto, J.~Rodr\'\i{}guez-Quintero and S.~Zafeiropoulos,
Phys. Rev. D \textbf{96}, no.9, 098501 (2017)
doi:10.1103/PhysRevD.96.098501
[arXiv:1704.02053 [hep-lat]].

\bibitem{Duarte:2017wte}
A.~G.~Duarte, O.~Oliveira and P.~J.~Silva,
Phys. Rev. D \textbf{96}, no.9, 098502 (2017)
doi:10.1103/PhysRevD.96.098502
[arXiv:1704.02864 [hep-lat]].

\bibitem{Krein:1990sf}
G.~Krein, C.~D.~Roberts and A.~G.~Williams,
Int. J. Mod. Phys. A \textbf{7}, 5607-5624 (1992)
doi:10.1142/S0217751X92002544

\bibitem{Bowman:2002bm}
P.~O.~Bowman, U.~M.~Heller and A.~G.~Williams,
Phys. Rev. D \textbf{66}, 014505 (2002)
doi:10.1103/PhysRevD.66.014505
[arXiv:hep-lat/0203001 [hep-lat]].

\bibitem{Boucaud:2003dx}
P.~Boucaud, F.~de Soto, J.~P.~Leroy, A.~Le Yaouanc, J.~Micheli, H.~Moutarde, O.~Pene and J.~Rodriguez-Quintero,
Phys. Lett. B \textbf{575}, 256-267 (2003)
doi:10.1016/j.physletb.2003.08.065
[arXiv:hep-lat/0307026 [hep-lat]].

\bibitem{Fischer:2005nf}
C.~S.~Fischer and M.~R.~Pennington,
Phys. Rev. D \textbf{73}, 034029 (2006)
doi:10.1103/PhysRevD.73.034029
[arXiv:hep-ph/0512233 [hep-ph]].

\bibitem{Furui:2006rx}
S.~Furui and H.~Nakajima,
Phys. Rev. D \textbf{73}, 094506 (2006)
doi:10.1103/PhysRevD.73.094506
[arXiv:hep-lat/0602027 [hep-lat]].

\bibitem{Furui:2006ks}
S.~Furui and H.~Nakajima,
Phys. Rev. D \textbf{73}, 074503 (2006)
doi:10.1103/PhysRevD.73.074503

\bibitem{Ayala:2012pb}
A.~Ayala, A.~Bashir, D.~Binosi, M.~Cristoforetti and J.~Rodriguez-Quintero,
Phys. Rev. D \textbf{86}, 074512 (2012)
doi:10.1103/PhysRevD.86.074512
[arXiv:1208.0795 [hep-ph]].

\bibitem{Burgio:2012ph}
G.~Burgio, M.~Schrock, H.~Reinhardt and M.~Quandt,
Phys. Rev. D \textbf{86}, 014506 (2012)
doi:10.1103/PhysRevD.86.014506
[arXiv:1204.0716 [hep-lat]].

\bibitem{Dorkin:2013rsa}
S.~M.~Dorkin, L.~P.~Kaptari, T.~Hilger and B.~Kampfer,
Phys. Rev. C \textbf{89}, 034005 (2014)
doi:10.1103/PhysRevC.89.034005
[arXiv:1312.2721 [hep-ph]].

\bibitem{Dorkin:2014lxa}
S.~M.~Dorkin, L.~P.~Kaptari and B.~K\"ampfer,
Phys. Rev. C \textbf{91}, no.5, 055201 (2015)
doi:10.1103/PhysRevC.91.055201
[arXiv:1412.3345 [hep-ph]].

\bibitem{Fu:2015tdu}
H.~F.~Fu and Q.~Wang,
Phys. Rev. D \textbf{93}, no.1, 014013 (2016)
doi:10.1103/PhysRevD.93.014013
[arXiv:1511.01587 [hep-ph]].

\bibitem{Windisch:2016iud}
A.~Windisch,
Phys. Rev. C \textbf{95}, no.4, 045204 (2017)
doi:10.1103/PhysRevC.95.045204
[arXiv:1612.06002 [hep-ph]].

\bibitem{Solis:2019fzm}
E.~L.~Solis, C.~S.~R.~Costa, V.~V.~Luiz and G.~Krein,
Few Body Syst. \textbf{60}, no.3, 49 (2019)
doi:10.1007/s00601-019-1517-9
[arXiv:1905.08710 [hep-ph]].

\bibitem{Lessa:2022wqc}
J.~R.~Lessa, F.~E.~Serna, B.~El-Bennich, A.~Bashir and O.~Oliveira,
[arXiv:2202.12313 [hep-ph]].

\bibitem{Oliveira:2018lln}
O.~Oliveira, P.~J.~Silva, J.~I.~Skullerud and A.~Sternbeck,
Phys. Rev. D \textbf{99}, no.9, 094506 (2019)
doi:10.1103/PhysRevD.99.094506
[arXiv:1809.02541 [hep-lat]].

\bibitem{Comitini:2021kxj}
G.~Comitini, D.~Rizzo, M.~Battello and F.~Siringo,
Phys. Rev. D \textbf{104}, no.7, 074020 (2021)
doi:10.1103/PhysRevD.104.074020
[arXiv:2108.00417 [hep-th]].

\bibitem{Virgili:2022wfx}
A.~Virgili, W.~Kamleh and D.~Leinweber,
[arXiv:2209.14864 [hep-lat]].



\bibitem{Maris:1994ux}
P.~Maris,
Phys. Rev. D \textbf{50}, 4189-4193 (1994)
doi:10.1103/PhysRevD.50.4189

\bibitem{Maris:1995ns}
P.~Maris,
Phys. Rev. D \textbf{52}, 6087-6097 (1995)
doi:10.1103/PhysRevD.52.6087
[arXiv:hep-ph/9508323 [hep-ph]].

\bibitem{Burden:1997ja}
C.~J.~Burden,
Phys. Rev. D \textbf{57}, 276-286 (1998)
doi:10.1103/PhysRevD.57.276
[arXiv:hep-ph/9702411 [hep-ph]].

\bibitem{Bhagwat:2003vw}
M.~S.~Bhagwat, M.~A.~Pichowsky, C.~D.~Roberts and P.~C.~Tandy,
Phys. Rev. C \textbf{68}, 015203 (2003)
doi:10.1103/PhysRevC.68.015203
[arXiv:nucl-th/0304003 [nucl-th]].


\bibitem{Frederico:2019noo}
T.~Frederico, D.~C.~Duarte, W.~de Paula, E.~Ydrefors, S.~Jia and P.~Maris,
[arXiv:1905.00703 [hep-ph]].

\bibitem{Hayashi:2021nnj}
Y.~Hayashi and K.~I.~Kondo,
Phys. Rev. D \textbf{103}, no.11, L111504 (2021)
doi:10.1103/PhysRevD.103.L111504
[arXiv:2103.14322 [hep-th]].

\bibitem{Hayashi:2021jju}
Y.~Hayashi and K.~I.~Kondo,
Phys. Rev. D \textbf{104}, no.7, 074024 (2021)
doi:10.1103/PhysRevD.104.074024
[arXiv:2105.07487 [hep-th]].

\bibitem{Sauli:2020dmx}
V.~Sauli,
[arXiv:2011.00536 [hep-lat]].


\bibitem{Falcao:2020vyr}
A.~F.~Falc\~ao, O.~Oliveira and P.~J.~Silva,
Phys. Rev. D \textbf{102}, no.11, 114518 (2020)
doi:10.1103/PhysRevD.102.114518
[arXiv:2008.02614 [hep-lat]].

\bibitem{Boito:2022rad}
D.~Boito, A.~Cucchieri, C.~Y.~London and T.~Mendes,
[arXiv:2210.10490 [hep-lat]].


\bibitem{Pomm73}
C.~Pommerenke, 
J. Math. Anal. Appl. \textbf{41}, 775  (1973) 

\bibitem{Baker75}
G.~A.~Baker, ``Essentials of Pad\'e approximants", Academic Press, U.S.A., (1975)

\bibitem{BaMo96}
G.~A.~Baker and P.~Graves-Morris, ``Encyclopedia of mathematics and its applications",
Cambridge University Press, Cambridge, U.K., (1996)

\bibitem{BeOr99}
C.~Bender and S.~Orszag, ``Advanced mathematical methods for scientists and engineers I:
asymptotic methods and perturbation theory", Springer, New York, U.S.A., (1999)

\bibitem{SanzCillero:2010mp}
J.~Sanz-Cillero,
[arXiv:1002.3512 [hep-ph]].

\bibitem{Queralt:2010sv}
P.~Masjuan Queralt,
[arXiv:1005.5683 [hep-ph]].


\bibitem{Math}
\url{https://www.wolfram.com/mathematica/}   









\bibitem{Li:2021wol}
S.~W.~Li, P.~Lowdon, O.~Oliveira and P.~J.~Silva,
Phys. Lett. B \textbf{823}, 136753 (2021)
doi:10.1016/j.physletb.2021.136753
[arXiv:2109.10942 [hep-th]].


\bibitem{Binosi:2019ecz}
D.~Binosi and R.~A.~Tripolt,
Phys. Lett. B \textbf{801}, 135171 (2020)
doi:10.1016/j.physletb.2019.135171
[arXiv:1904.08172 [hep-ph]].


\bibitem{Zwanziger:1989mf}
D.~Zwanziger,
Nucl. Phys. B \textbf{323}, 513-544 (1989)
doi:10.1016/0550-3213(89)90122-3

\bibitem{Dudal:2008sp}
D.~Dudal, J.~A.~Gracey, S.~P.~Sorella, N.~Vandersickel and H.~Verschelde,
Phys. Rev. D \textbf{78}, 065047 (2008)
doi:10.1103/PhysRevD.78.065047
[arXiv:0806.4348 [hep-th]].

\bibitem{Dudal:2007cw}
D.~Dudal, S.~P.~Sorella, N.~Vandersickel and H.~Verschelde,
Phys. Rev. D \textbf{77}, 071501 (2008)
doi:10.1103/PhysRevD.77.071501
[arXiv:0711.4496 [hep-th]].




\bibitem{Oliveira:2012eh}
O.~Oliveira and P.~J.~Silva,
Phys. Rev. D \textbf{86}, 114513 (2012)
doi:10.1103/PhysRevD.86.114513
[arXiv:1207.3029 [hep-lat]].


\bibitem{Cucchieri:2016jwg}
A.~Cucchieri, D.~Dudal, T.~Mendes and N.~Vandersickel,
Phys. Rev. D \textbf{93}, no.9, 094513 (2016)
doi:10.1103/PhysRevD.93.094513
[arXiv:1602.01646 [hep-lat]].


\bibitem{Yndurain:2006amm}
F.~J.~Yndurain,
``The Theory of Quark and Gluon Interactions,''
doi:10.1007/3-540-33210-3


\bibitem{Vandersickel:2012tz}
N.~Vandersickel and D.~Zwanziger,
Phys. Rept. \textbf{520}, 175-251 (2012)
doi:10.1016/j.physrep.2012.07.003
[arXiv:1202.1491 [hep-th]].




\bibitem{Aguilar:2017dco}
A.~C.~Aguilar, D.~Binosi, C.~T.~Figueiredo and J.~Papavassiliou,
Eur. Phys. J. C \textbf{78}, no.3, 181 (2018)
doi:10.1140/epjc/s10052-018-5679-2
[arXiv:1712.06926 [hep-ph]].



\bibitem{Delbourgo:1977jc}
R.~Delbourgo and P.~C.~West,
J. Phys. A \textbf{10}, 1049 (1977)
doi:10.1088/0305-4470/10/6/024

\end{thebibliography}
\end{document}